\documentstyle[aps,pre,epsf,eqsecnum]{revtex}

\begin{document}

\title{Statistics of Self-Crossings and Avoided Crossings of Periodic Orbits
in the Hadamard-Gutzwiller Model}

\author{Peter A. Braun, Stefan Heusler, Sebastian M\"uller, and Fritz Haake}
\address{Fachbereich Physik, Universit\"at Essen,
45\,117 Essen, Germany }
\date{\today}
\maketitle
\begin{abstract}

Employing symbolic dynamics for geodesic motion on the tesselated pseudosphere, the so-called
Hadamard-Gutzwiller model, we construct extremely long periodic orbits without compromising
accuracy. We establish criteria for such long orbits to behave ergodically and to yield 
reliable statistics for self-crossings and avoided crossings.  Self-encounters of periodic 
orbits are reflected in certain patterns within symbol sequences, and these allow for analytic 
treatment of the crossing statistics. In particular, the distributions of crossing angles and
avoided-crossing widths thus come out as related by analytic
continuation. Moreover, the action difference for Sieber-Richter pairs of orbits (one
orbit has a self-crossing which the other narrowly avoids and otherwise the orbits look very
nearly the same) results to all orders in the crossing angle. These findings may be helpful for 
extending the work of Sieber and Richter towards a fuller understanding of the classical basis
of quantum spectral fluctuations.

\end{abstract}

\section{Introduction}

\vspace{1cm} Billiards on surfaces of negative curvature were
first investigated by J. Hadamard \cite{1898}. The case of
constant negative curvature, known as the pseudosphere, has
enjoyed considerable popularity since Gutzwiller's realization
\cite{Gutzibill} of its potential as a paradigm of quantum chaos.
Complete hyperbolicity, the availability of symbolic dynamics, the
equality of the Lyapunov exponents of all periodic orbits, and the
validity of Selberg's trace formula are among the attractive
features of that system. Useful introductions can be found in
Refs.~\cite{Gutzibuch,BalVor,austei0,geobo}. For a tesselation by maximally
desymmetrized octagons (see below) Aurich and Steiner found the spectral
fluctuations of the quantum energy spectrum faithful to the
Gaussian orthogonal ensemble (GOE) of random-matrix theory
\cite{austei}, as illustrated in Fig.~1 for the so-called form
factor, the Fourier transform of the energy dependent two-point
correlator of the density of levels.

One of the urgent problems in quantum chaology is to understand
the rather universal validity of random-matrix type spectral
fluctuations for chaotic dynamical systems, nowadays known as the
Bohigas-Giannoni-Schmit conjecture \cite{BGS}. An important first
step was made by Berry \cite{Berry} on the basis of Gutzwiller's
trace formula \cite{tracef} which expresses the oscillatory part of the density of energy
levels as a sum over periodic orbits, $d_{{\rm osc}}(E)=\mbox{Re}\sum_\gamma A_\gamma
{\rm e}^{{\rm i} S_\gamma/\hbar}$, with $S_\gamma$ the action and
$A_\gamma$ a classical stability amplitude of the periodic orbit
$\gamma$. Berry realized that the double sum over periodic orbits
in the form factor, which involves the building block $A_\gamma
A_{\gamma'}^*{\rm e}^{{\rm i}(S_\gamma-S_{\gamma'})/\hbar}$ in each summand, must draw an important
contribution from the diagonal terms $\gamma=\gamma'$ since pairs
of orbits with action differences larger than Planck's unit
$\hbar$ can be expected to interfere destructively and thus to
cancel in the form factor. For time reversal invariant systems
each orbit $\gamma$ and its time reverse $\gamma^{\rm TR}$ have
equal action such that Berry's ``diagonal approximation''
generalizes to include pairs of mutually time reversed orbits and
then gives the time dependent form factor as
$K(\tau)=2\tau+\ldots$ where $\tau$ is the time in units of the so
called Heisenberg time, which is given in terms of the mean level
density $\bar{d}(E)$ as $T_{\rm H}(E)=2\pi\hbar \bar{d}(E)$.

Recently, Sieber and Richter \cite{gedisieb,siebrich} employed the
pseudosphere in their pioneering move beyond the diagonal
approximation; they found a one-parameter family of orbit pairs
within which the action difference can be steered to zero. One
orbit within each ``Sieber-Richter pair'' undergoes a small-angle
self-crossing which the partner orbit narrowly avoids. The form
factor receives the contribution $K_{\rm off}^1(\tau)=-2\tau^2$
from the family. Together with Berry's $K_{\rm diag}(\tau)=2\tau$
we thus have a semiclassical understanding of at least the first
two terms in the Taylor series of the random-matrix form factor
\cite{haake} $K_{\rm GOE}(\tau)= 2\tau-\tau\ln(1+2\tau)$. The
search is now on for further families of orbit pairs which might
yield the higher-order terms of the expansion. At the same time,
one would like and will eventually have to go beyond the
pseudosphere, in order to find the conditions for universal
behavior; interesting first steps have been taken for quantum
graphs \cite{qugra} and other billiards \cite{Muell}.

In the present paper we remain with the pseudosphere for a
thorough investigation of action correlations which we feel
necessary as a basis for further progress towards an understanding
of the random-matrix like spectral fluctuations in this
prototypical dynamical system (and beyond). We shall make
extensive use of symbolic dynamics in (i) establishing a certain
local character of the relation between an orbit and its symbol
sequence (only a section of the symbol sequence is necessary to
determine the associated section of the orbit), (ii) constructing
very long periodic orbits (up to hundreds of thousands of symbols)
with full control of accuracy, (iii) identifying the ergodic ones
among them and establishing a simple and instructive rederivation
of Huber's exponential-proliferation law, (iv) expressing the
angle $\epsilon$ of a self-crossing and the closest-approach
distance $\delta$ in the partner orbit of a Sieber-Richter pair in
terms of the M\"obius-transformation matrices associated with the
loops joined in a crossing, (v) revealing a useful
analytic-continuation kinship of $\epsilon$ and $\delta$, (vi)
explicitly relating the joint density $P(\epsilon,l|L)$ for
crossing angles and loop lengths $l$ in orbits of total length $L$
to the associated density  $P^a(\delta,l|L)$ for avoided
crossings, and (vii) constructing an expression for the action
difference of a Sieber-Richter pair valid to all orders in
$\epsilon$.

Some of our findings are based on (overwhelming and exceptionless)
numerical evidence based on large numbers of long periodic orbits
and thus call for mathematical substantiation.

Even though we strictly confine ourselves to the pseudosphere we
expect (and have indeed begun to check) generalizability to other
systems for which symbolic dynamics is available.

\begin{figure}
\begin{center}
\leavevmode \epsfxsize=0.35 \textwidth \epsffile{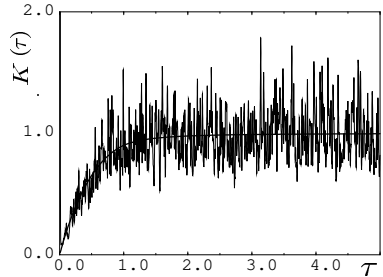}
\end{center}
\caption{ Form factor $K(\tau)$ for an
asymmetric octagon based on energy levels determined by Aurich and
Steiner \protect\cite{austei}, after averaging over a time window $\Delta\tau=0.01$. Smooth line for GOE.} \label{noein}
\end{figure}

\section{The Hadamard-Gutzwiller model}\label{Hadamard}

The Hadamard-Gutzwiller model \cite{BalVor,austei} is a
two-dimensional billiard on the so called Poincar\'{e} disc, i.e.
  the unit disc $x^2 + y^2 = |z|^2 \leq 1$ endowed with the metric
\begin{equation}
\label{metric} d s^2 = 4 \frac{d x^2 + d y^2}{(1 - x^2 - y^2)^2}
=4 \frac{ d z d z^*} {(1 - |z|^2)^2}\,.
\end{equation}
The distance $d(z_1, z_2)$ between two points $z_1, z_2$, measured
along the unique geodesic connecting them, reads
\begin{equation}
\label{hypdist}
\cosh d(z_1, z_2) = 1 + \frac{2 |z_1 -
z_2|^2}{(1 - |z_1|^2)(1- |z_2|^2)}\,.
\end{equation}

 The geodesics are
circles intersecting the boundary at right angles. The total length of
the geodesics between its two crossing points with the disc boundary is
infinite signaling non-compactness.  Free motion on that space is
completely hyperbolic. All trajectories (geodesics) have the same
Lyapunov exponent, $\lambda = 1$, and that fact makes for great
simplifications as compared to usual billiards in flat space. The
Poincar\'{e} disc can be regarded as the stereographic projection of
the pseudosphere, the surface of constant negative curvature.

The symmetry of the Poincar\'{e} disc is described by the
non-compact Lorentz group $SU(1,1)$; we shall only be interested in its hyperbolic elements which are  matrices with the structure
\begin{equation}
{\cal M}=\left ( \begin{array}{cc} \alpha & \beta \\ \beta^* &
\alpha^*
\end{array} \right)\,,\quad |\alpha|^2 - |\beta|^2 = 1\,,\quad
|{\rm Re}\, \alpha| > 1\,. \label{su11matrix}
\end{equation}
Their action on  points of the complex plane is defined as the
M\"{o}bius transformation
\begin{equation}
z'={\cal M}(z) := \frac{\alpha z + \beta}{\beta^* z + \alpha^*}
\,.
\end{equation}
Inner points of the Poincar\'{e} disc are transformed to 
inner points, and  boundary points  $|z|=1$ to boundary points.
Two boundary points  $z_S, z_U$  remain invariant, $z_S={\cal M}
(z_S)$ and similar for $z_U$; the indices signaling ``stable''
and ``unstable''. Using ${\rm det} \ {\cal M}=1$ we find

\begin{equation}
\label{fp} z_{S, U} = \frac{1}{2 \beta^*} \Big( \alpha - \alpha^*
\pm \sqrt{(\alpha + \alpha^*)^2 - 4}\Big)
\end{equation}
where $S$ refers to the $+$ sign, and $U$ to the $-$ sign in the
case ${\rm Re}\,\alpha>1$; if ${\rm Re}\,\alpha<-1$ the opposite
sign assignment applies. Repeated application of the
transformation ${\cal M}$ on any point $z$ leads to the stable
fixed point, ${\cal M}^k(z) \rightarrow z_S$ for $k=1,2,3,\ldots$,
while iteration of the inverse map leads to the unstable fixed
point, ${\cal M}^{-k}(z) \rightarrow z_U$.

 The geodesic passing through $z_S$ and $z_U$ (to be called the ``own''
geodesic of the matrix $ {\cal M}$) is invariant with respect to
${\cal M}$: if $z $  belongs to this geodesic then so does $z'=
{\cal M}(z)$. The distance $d(z,{\cal M}(z))$ is the same  for all points of
the geodesic and equals

\begin{equation}
\label{1.1} L=d(z,z')=2 \mbox{arccosh} \frac {|\mbox{ Tr }{\cal
M}|}{2}\,.
\end{equation}
The inverse ${\cal M}^{-1}$  of a matrix ${\cal M}$ is obtained by the replacement 
$\alpha\to \alpha ^*,\beta\to -\beta$ in (\ref{su11matrix}). These two matrices have
their stable and unstable points  interchanged. Consequently,
a matrix and its inverse  shift points along their common ``own''
geodesic by the same distance but in opposite directions.

To obtain a model with compact configuration space the Poincar\'{e}
disc is tesselated with tiles of equal area and shape. Each type of 
tesselation is connected
with a particular discrete infinite subgroup  $\Gamma$ of
$SU(1,1)$ such that by acting on all inner points of a tile by some fixed matrix 
${\cal W}\ne {\bf 1}$ we get all inner points of some other tile; the boundaries  are
mapped to boundaries of the same or other tiles. Using all ${\cal
W}\in\Gamma$ we get all tiles and restore the whole disc from a single initial tile.   A
particular tile is distinguished by containing the origin $z=0$
and is called the fundamental domain. In fact,
different tiles are identified by identifying each point $z$ of the Poincar\'e disc 
with all its images ${\cal W}(z)$, and so a compact configuration space is indeed arrived at.

\begin{figure}
\begin{center}
\leavevmode \epsfxsize=0.35 \textwidth \epsffile{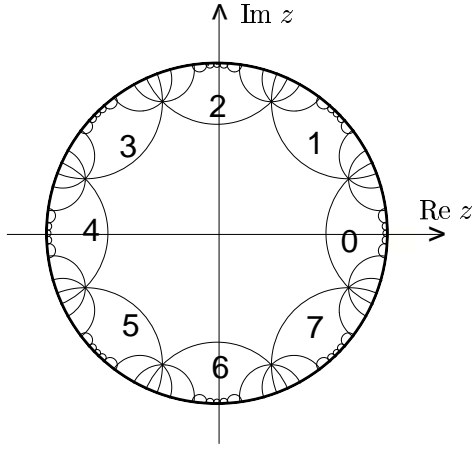}
\end{center}
\caption{Poincar\'{e} disc and its tesselation with regular octagons.} \label{nozwei}
\end{figure}

In the present paper we  exclusively work with  octagonal tiles which are all identified with 
the fundamental domain. The opposite  sides of the latter are glued together such that we arrive at  a Riemann surface of genus 2, i.e. a two-hole
doughnut. Matrices ${\cal W}$ 
giving rise to octagonal tiles
are arbitrary products of four elementary group elements  $l_0,
l_1, l_2, l_3 \in SU(1, 1)$ and their inverses. For simplicity, we
shall mostly consider tesselation with ``regular octagons '' (Fig.~\ref{nozwei}) which
have the highest possible symmetry; the pertinent elementary matrices  $l_k,\,
k=0,1,2,3,$ are
\begin{equation}
l_{k}=\left (
\begin{array}{cc}
1+\sqrt{2} & \sqrt{2+2\sqrt{2}}\,{\rm e}^{{\rm i}\pi k/4} \\
\sqrt{2+2\sqrt{2}}\,{\rm e}^{-{\rm i}\pi k/4} & 1+\sqrt{2}
\end{array}
\right) \;\label{elem}
\end{equation}
and their inverses. The inverse of $l_k$ is easily checked to be $l_{k+4}$. Since the
index $k$ describes a phase in the off-diagonal matrix elements we
conclude
\begin{equation}\label{invell}
l_k^{-1}=l_{k+4} = l_{(k+4){\rm\mbox{ }mod\mbox{ }}8}\equiv l_{\bar{k}} \,,
\end{equation}
where we have introduced $\bar{k}= (k+4){\rm\mbox{ }mod\mbox{ }}8$. The full set of elementary
matrices is thus $l_k,\, k=0,\ldots,7$.

The sides of the boundary of the fundamental domain can now be
labeled $0,1,\ldots,7$ and constructed as follows. Tesselation
identifies
 opposite sides of the regular octagon  as 0 $\equiv$ 4, 1 $\equiv$ 5, 2 $\equiv$ 6, and 3
$\equiv$ 7. Points on side $k+4$ are thus images of their mirror
symmetric counterparts on side $k$; the group element responsible
for such imaging is $l_{k+4}$. In particular, the points on side 0
solve the quadratic equation $l_4(z)=-z^*$ and form a circle of
radius $\left(2^{-1/2}-2^{-1}\right)^{1/2}$, which is in fact a
geodesic. Side $k$ is obtained by a rotation of side 0 by
$k\pi/8$.

Elementary matrices other than $l_{k+4}$ lead from points  on side $k$
of the fundamental domain to boundary sides in the next generation of
tiles, which may be called  "higher Brillouin zones", using an analogy
with periodic lattices.  Obviously, there are eight Brillouin zones
neighboring to the fundamental domain. In all other Brillouin zones of
all generations, opposite sides  are still identified. All
Brillouin zones have  the same octagonal shape and the same
area as the fundamental domain; their visual difference in size and
form is  caused by the metric (\ref{metric}).

In what follows, we shall often call the elementary matrices $l_k$
"letters", and their products $l_{j_1} l_{j_2}l_{j_3}\ldots
l_{j_n}$ "words". As a convenient shorthand for a word we shall
also employ just the string of indices of the letters, i.e. ${\cal
W} = l_{j_1} l_{j_2} l_{j_3} \ldots l_{j_n} \equiv (j_1, j_2,
j_3,\ldots j_n)$.
\begin{figure}
\begin{center}
\leavevmode \epsfxsize=0.3 \textwidth \epsffile{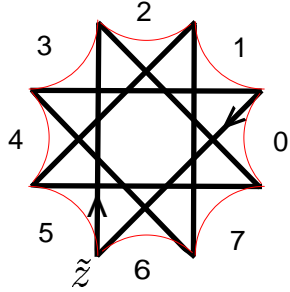}
\end{center}
\caption{ Corner points identified by the group identity (\ref{ide}); starting from ${\tilde z}$ 
and following the arrow, the identity can be read off. }
\label{gi}
\end{figure}

As already mentioned, the identification of points $z\equiv{\cal
W}(z)$ implies gluing together opposite sides of the octagon; the
result is a Riemann surface of genus $2$. On that surface, the
eight corner points of the octagon in the unit disc coincide. The
identification of the corner points is shown in Fig.~\ref{gi} .
Starting from any corner, (for instance ${\tilde z}$ in the
figure), taking into account that elementary matrices map opposite
points of the boundaries of the fundamental domain onto each
other, we go through all corners and come back after eight steps,
and read off $l_5  l_0  l_3  l_6  l_1 l_4 l_7 l_2({\tilde z}) =
{\tilde z} $. Due to the fact that the fixed-point equation ${\cal
W}(z) = z$ for ${\cal W} \neq {1\,0\choose 0\,1}$ has unimodular
solutions $|z_{S, U}| = 1$, we conclude that the matrix product
$l_5 l_0  l_3  l_6  l_1 l_4 l_7 l_2$ must be the identity, since
the corner point ${\tilde z}$ obviously fulfills ${\tilde z} \neq
1$ . In fact, using the explicit form (\ref{elem}) of the $l_k$,
one checks that the foregoing product of eight matrices is the
$2\times2$ unit matrix,

\begin{equation}
\label{ide} l_5  l_0  l_3  l_6  l_1 l_4 l_7 l_2 \equiv (5, 0, 3,
6, 1, 4, 7, 2) ={1\,0\choose 0\,1} = \mathbf{1} \, .
\end{equation}

Any geodesic in the tesselated unit disc can be folded into the
fundamental domain where it will  look like a sequence of disjoint
circular segments starting and ending on the boundaries of the
octagon. Of course, when the fundamental tile is regarded as a
surface  of genus 2 the circular sections in question no longer
appear as disjoint: A trajectory leaving the fundamental domain
through one of the eight sides of the octagon and reentering on
the opposite side appears as a smooth curve on the genus 2
surface. It is only after representing the whole unit disc by a
single tile with opposite sides glued together (or, equivalently
by a surface of genus 2) that a geodesic is capable of
self-crossings.

Consider now inertial motion along the ``own'' geodesic of a matrix
${\cal W}\in\Gamma$. Any point $z$ of the geodesic and its image ${\cal
W}(z)$ ( which we know to belong to the same geodesic) are identified
by the tesselation. Geodesically moving from $z$ to ${\cal W}(z)$ we are
in fact traversing a periodic orbit associated with ${\cal W}$, and the 
length of that periodic orbit is given by (\ref{1.1});
the time reversed orbit is similarly associated with the matrix
${\cal W}^{-1}$. Since each ${\cal W}\in\Gamma$ can be written as a
product of elementary matrices ${\cal W}=l_{j_1}l_{j_2}\ldots
l_{j_n}$ and encoded by the symbolic word
$\{j_1,l_2,\ldots l_n\}$ we have in fact symbolic dynamics at our disposal as
a  tool for investigating of periodic orbits.\footnote{Note
that we take ``symbol'' and ``letter'' as synonyms; neither do we
distinguish ``symbol sequence'', ``word'', ``equivalence class of
words'', and ``orbit'', unless necessary; whenever a notational
distinction seems helpful we use $W,A,\ldots$ for words in the
sense of symbol sequences and ${\cal W,A,}\ldots$ for the
associated M\"obius-transformation matrices.}

All matrices from an equivalence class ${\cal Z}{\cal
W}{\cal Z}^{-1}$, where ${\cal Z}$ is any matrix from $ \Gamma$,
have their ``own'' geodesics  identified by tesselation;
there is thus one and one only periodic orbit per equivalence
class in $\Gamma$. An infinite number of words pertaining to the same 
equivalence class and thus referring to
the same periodic orbit can be obtained  from one representative
word ${\cal W}=(j_1, j_2, j_3, ...j_r)$; this is done by cyclic
permutations of the letters, by similarity transforms (replacing
${\cal W}$ by the longer word ${\cal Z}{\cal W}{\cal Z}^{\rm TR}$ where
${\cal Z}$ is another word, and its associate ${\cal Z}^{\rm TR}={\cal
Z}^{-1}$ stands for the matrix inverse to the one represented by ${\cal
Z}$; the superscript ``TR'' is read as ``time reversed''), and by
using the group identity (\ref{ide}) in all its various forms.  A code
of the time reversed periodic orbit is produced from the original ${\cal W}$
by writing its symbols $j_i$ in the opposite order and replacing each
$j_i$ by $\bar{j}_i=j_i+4 \mbox{ mod } 8$.

Like any geodesic the periodic orbit can be folded into the
fundamental domain where it will consist of a finite number $n$ of
disjoined circular segments. After that a distinguished $n$-letter
word can be introduced for the orbit which is simply the sequence
of the ``landing'' sides of the orbit segments. (The ``launching''
side is always opposite to  the ``landing'' side of the
preceding segment and is not admitted to the symbolic code.) This encoding
contains the least possible number of symbols among all members of
the  equivalence class, and is unique up to cyclic permutations. All 
$n$ ``own'' geodesics of a matrix ${\cal W}$ encoded by the distinguished 
word and its cyclic permutations, cross the fundamental domain.

\begin{figure}
\begin{center}
\leavevmode \epsfxsize=0.25 \textwidth \epsffile{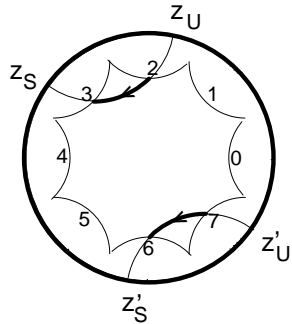}
\end{center}
\caption{ Two  equivalent geodesics depicting 
the periodic orbit $(3,6)$; they connect the  fixed points
$z_U,z_S$  of  $l_3 l_6$ and  $z_U',z_S'$ of  $l_6l_3$. 
The primitive orbit inside the octagon (bold parts of the geodesics) 
has two segments.} \label{bz}
\end{figure}

As an example of how to construct an orbit from its symbol
sequence we consider the word ${\cal W}=(3,6)$. Its ``own''
geodesic  can be found calculating the matrix product $l_3l_6$ and
geodesically joining the respective fixed points (Eq.(\ref{fp})), the curve 
$z_Sz_U$ in Fig.~\ref{bz}. The full orbit within the fundamental domain could  be
obtained by folding the geodesic $z_Sz_U$ back into that domain. However, it is 
easier to find  the``own'' geodesic of the cyclicly permuted matrix
${\cal W}'=l_6l_3$ and joining its fixed points $z_{S,U}'$.  The orbit
in the fundamental domain is now given by the ``inner parts'' of
the two geodesics (bold intervals of $z_Sz_U$ and $z_S'z_U'$ in
Fig.~\ref{bz}   ). The ``non-primed'' segment of the orbit starts
on side 2 and ends on side 3, while the ``primed'' one starts on
7 and ends on 6. The time reversed orbit would have the
launching and landing sides interchanged such that its word would be
$(2,7)$.

Tesselation with less symmetric octagons, also
corresponding to a genus 2 Riemann surface, can be implemented
with four elementary matrices $l_j \in SU(1, 1)$ and their
inverses $l_{\bar j}$, chosen such that all eight matrices obey
the group identity (\ref{ide}).  Interestingly, completely
desymmetrized genus 2 surfaces do not exist: the inversion
symmetry under $z \to -z$ can only be destroyed for $g \geq 3$
\cite{Abresch}.

\section{Construction of long periodic orbits}\label{longorb}

We here propose a new method for constructing very long periodic
orbits. Let us consider a many-letter word ${\cal W}=(j_1, j_2,
j_3,...j_n)$ and its matrix $l_{j_1} l_{j_2} l_{j_3}...l_{j_n}$.
In order to explicitly construct the associated periodic orbit we
can proceed stepwise, so as to determine each of the $n$ circular
segments  separately. Each step uses
only part of the word ${\cal W}$ (the symbol assigned to a segment and 
its near neighbors); the necessary length of that
part is determined by the required accuracy and not at all by the
length of ${\cal W}$.

In the preceding section, we discussed the connection between the fixed points $z_S,
z_U$ of the M\"obius transformation associated with ${\cal W} =
l_{j_1} l_{j_2} l_{j_3}...l_{j_n}$ and the periodic orbit. We would also like to recall that the two fixed points
determine the circular segment associated with the first letter
$l_{j_1}$ of the word; to find the next circular segment one has
to determine the two fixed points for the equivalent word obtained
by cyclically permuting $l_{j_1}$ to the right end. This is how
all $n$ circular segments of the orbit in question can be found,
one after the other.

It is convenient to first determine the stable fixed point for the
first circular segment. To that end, we consider the sequence of
matrices $\{ {\cal W}_1=l_{j_1}, {\cal W}_2=l_{j_1}l_{j_2}, {\cal
W}_3=l_{j_1}l_{j_2}l_{j_3},\ldots \}$ which are obtained by
truncating ${\cal W}$. For each matrix in the sequence we solve
the quadratic fixed-point equation. The sequence of unstable fixed
points thus obtained behaves erratically. The sequence of stable
fixed points, however, converges rapidly, in fact with
exponentially growing accuracy. The
limiting point itself is none other than the stable fixed point
$z_S$ of the whole word beginning with $l_{j_1}$.

The convergence of the series of stable fixed points of ${\cal W}_k$  can be ascertained analytically. To that end we first observe that the matrix
${\cal W}_k$ has its elements grow exponentially with the length $L$ of
its corresponding periodic orbit. This is obvious from the
relation (\ref{1.1}) of the length of an orbit to the trace of
its matrix, $|{\rm Tr} \ {\cal W}_k| = 2 \cosh(L_k/2) \approx {\rm e}^{L_k/2}$,
and from the unimodularity of the determinant, ${\rm det}\  {\cal
W}_k =\alpha \alpha^* - \beta\beta^*  = 1$. While under other
circumstances such exponential growth gives rise to inaccuracy
growing out of control, we can now rejoice in the growth working
in our favor when determining the stable fixed point of ${\cal
W}$: For large matrix elements, Eq. (\ref{fp}) simplifies to

\begin{equation}
\label{qwq} z_S \approx \frac{\alpha}{\beta^*} \  \ \ \ \ \ \ \
z_U \approx -\frac{\alpha^*}{\beta^*} \,.
\end{equation}
When proceeding to ${\cal W}_{k+1}$ by including the letter
$l_{j_{k+1}} = \left({l_{11} \atop l_{12}^*}{l_{12}\atop
l_{11}^*}\right)$ and comparing the stable fixed points
$z_S^k,\,z_S^{k+1}$ of ${\cal W}_k =
\left({\alpha\atop\beta^*}{\beta\atop\alpha^*}\right) ,\,{\cal
W}_{k+1} = {\cal W}_k l_{j_{k+1}}$, we find for their difference,
using the approximation (\ref{qwq}) and the unimodularity of $\det
{\cal W}_k$,

\begin{equation}
z_S^k - z_S^{k+1} = \frac{l_{12}^*}{\beta^* (\beta^* l_{11} +
\alpha^* l_{12}^*)}\,.
\end{equation}
Since the elements $l_{ij}$ are of order unity, the difference
$z_S^k - z_S^{k+1}$ is of order ${\rm e}^{-L_k}$, i.e. small
once the word ${\cal W}_k$ is long. Exponentially fast convergence
of the sequence $z_S^k$ is thus obvious.

We can now turn to the task of determining the unstable fixed point of
$\cal W$. As explained in the last section, the time reversed orbit is
obtained using the reversed M\"obius transformation matrix ${\cal
W^{\rm TR}}={\cal W}^{-1}$, which has stable and unstable fixed
points interchanged relative to ${\cal W}$. Consequently, we get
the unstable fixed point of the circular segment of ${\cal W}$
associated with the letter $l_{j_1}$ as the limit of the sequence
of stable fixed points of the matrices $\{
l_{j_{n}}^{-1},l_{j_n}^{-1}l_{j_{n-1}}^{-1},l_{j_{n}}^{-1}
l_{j_{n-1}}^{-1} l_{j_{n-2}}^{ -1} ,\ldots \}$; these matrices are
truncations of ${\cal W}^{\rm TR}$. The sequence of stable fixed
points of these truncations converges as rapidly as the one considered before and yields the unstable fixed point $z_U$ for the first circular
segment of ${\cal W}$.

With both $z_U$ and $z_S$ determined to the
desired accuracy for the segment of the orbit corresponding to
$l_{j_1}$, one repeats the procedure for the subsequent circular
segment, i.e. the one corresponding to $l_{j_2}$.

As an obvious property of our method, we find that the
computational effort needed to construct a periodic orbit with
$n\gg 1$ symbols grows only linearly with $n$. More importantly,
we only have to do local work for local properties: Each of the $n$ circular segments of the orbit inside the
fundamental domain is found with a number of operations
independent of $n$; for each segment one has to calculate a product of a
moderate number of elementary matrices (about 18 for a 15-digit
accuracy). As we proceed segmentwise, no inaccuracy can accumulate. There is thus practically no limit on the length of the
accessible orbits. For example, we had no difficulty in
calculating a periodic orbit corresponding to a sequence of $10^5$
randomly selected symbols.

It may be useful to again comment on the fact that any periodic orbit is, in principle, determined by any one of its circular segments within the fundamental domain. One just has to take the full geodesic running between the fixed points $z_U,z_S$, to which the segment belongs, and fold that geodesi
c into the fundamental domain. However, such folding is numerically unstable and  cannot be implemented with finite precision for very long orbits.

Inasmuch as we work with fixed points of M\"obius transformations,
our method seems to be restricted to the Hadamard-Gutzwiller
model. However, underlying all technicalities is a locality in the
relationship between orbits and symbolic words, and that locality
does in fact make our method generalizable, as will be explained
in a separate publication.

\section{Pruning}

The locality of the word-orbit relationship makes our method a promising  pruning tool. Pruning generally means
recognizing and deleting symbolic words not corresponding to physically
realizable orbits \cite{Cvita}. In the Hadamard-Gutzwiller model
every word ${\cal W}$ does correspond to an allowed orbit. On the other
hand, there are infinitely many equivalent words which must be counted
only once whenever sums over orbits are to be taken, as for instance in
Selberg's trace formula.  The particular word we are interested in is
simply the ordered list of ``landing sides'' of the regular octagon for the succession of circular segments of the orbit (Section \ref{Hadamard}). The task  of selecting this
distinguished word can be regarded as a variant of pruning.

A first step is to discard  words which can be shortened,
considering that the distinguished representation contains the least
possible number of symbols; in particular, the word must not contain
pairs of symbols $k {\bar k}$ with ${\bar k} = (k + 4){\rm\mbox{ }mod\mbox{ }}8$.
The group identity (\ref{ide})  is yet another source of ``badness'':
Whenever it allows to replace a word by an equivalent shorter one, only
the latter needs to be further scrutinized.

Real problems are due to the fact that the group identity   (\ref{ide})
allows to write some  four-letter parts of words in reversed order, e.g.
$(0, 5, 2, 7) = (7, 2, 5, 0)$. The average number of such reversible
4-sequences in a word with $n$ random symbols is estimated  in Appendix A as
$0.0043n$. Each of the 4-sequences must have a uniquely defined
direction in the distinguished word; it is not
known {\it a priori}, however, which direction is the ``correct'' one.  The
suitable word has thus to be selected among $2^{0.0043n}$ candidates.
For, say, $n\sim 10^5$  no hope can be set on any
trial-and-error procedure like  building the orbits corresponding to
the equivalent words one by one  and discarding those not lying inside the
fundamental domain.

With our algorithm we had no problem finding the correct
direction of the $4$-letter sequences just mentioned. Whenever at
some stage our method produced an arc which did not cross the fundamental
domain, the letter for that arc turned up within a reversible
$4$-sequence. Upon reversing that particular $4$-sequence (out of
the thousands present in the code!) the arc always returned to the
fundamental domain, and construction of the periodic orbit could
be continued. Pruning thus becomes a local problem, which can be
attacked efficiently.

\section{Ergodic periodic orbits}\label{ergorb}

It is well known that the geodesic flow in the Hadamard-Gutzwiller
model is ergodic \cite{BalVor} such that almost all trajectories
cover the phase space homogeneously. Introducing limited
phase-space resolution  we can extend the notion of ergodicity to
periodic orbits: Almost all sufficiently long periodic orbits
cover the coarse-grained phase space uniformly.

\begin{figure}
\begin{center}
\leavevmode \epsfxsize=0.95 \textwidth \epsffile{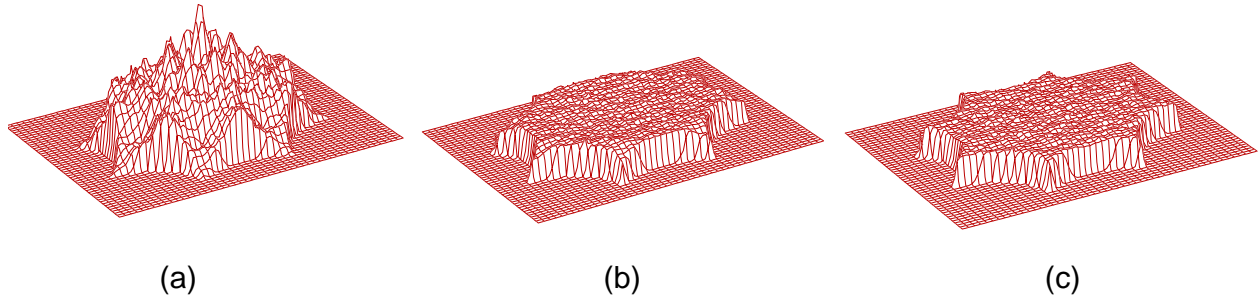}
\end{center}
\caption{Coverage of fundamental domain by three periodic 100000-symbol orbits: 
(a) Random symbol sequence yields extremely inhomogeneous distribution. (b) Account of two-step correlations gives almost ergodic distribution, with small deviations only near octagon corners. (c) Symbol sequence imported from  non-periodic trajectory gives uniform density.}
 \label{cake}
\end{figure}

The overwhelming prevalence of ergodic periodic
orbits does not imply that a {\it random} sequence of {\it uncorrelated} symbols must yield an ergodic orbit. In fact, practically every periodic
orbit so constructed is extremely non-ergodic, as illustrated in
Fig.~\ref{cake}(a) by the configuration-space density for 100000 randomly chosen symbols; the grossly non-uniform density thus incurred is anything but ergodic.

\begin{figure}
\begin{center}
\leavevmode \epsfxsize=0.6 \textwidth \epsffile{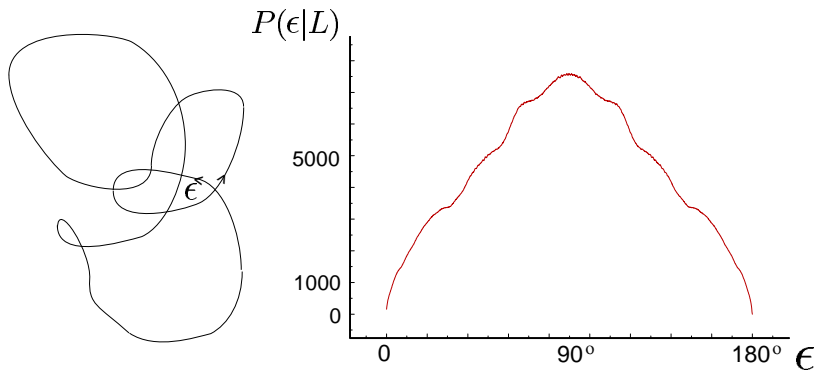}
\end{center}
\caption{Definition of crossing angle $\epsilon$; strongly non-ergodic density $P(\epsilon | L)$ for orbit with $10^5$ random symbols.}
\label{schnee}
\end{figure}

In configuration space, a long periodic orbit intersects itself many
times, and the distribution of self-crossing angles $\epsilon$
provides another sensitive test for ergodicity. We
shall consider it in some detail because of its role in the
Sieber-Richter theory. The angle $ \epsilon $, defined to lie in
the interval $ 0 \le \epsilon  < \pi $, is complementary to the
angle between the velocities at the  crossing; see
Fig.~\ref{schnee}. The  number of self-crossings  with crossing angles
in the interval $(\epsilon, \epsilon + d \epsilon)$ in periodic orbits of length
$L$ yields a density $P(\epsilon | L)$. Since no direction is distinguished the
probability that an element $d {\vec l}_1$ of the
orbit intersects with the element $d {\vec l}_2$ is  given by the
geometric projection $P(\epsilon | L) \propto |d{\vec l}_1 \times
d {\vec l}_2| \propto \sin \epsilon$.
Comparing that prediction with the plot of
$P(\epsilon | L)$, numerically obtained for our periodic orbit
with 100000 randomly chosen symbols, we encounter blatant disagreement, despite the
enormous length of the orbit. In particular, for small $\epsilon$,
the number $P(\epsilon | L)$ of self-crossings decreases like
$\propto \epsilon^{2/3}$ rather than $\propto \epsilon$.

It is important to stress that the non-ergodic patterns in
Figs.~\ref{cake}(a) and \ref{schnee} are not due to an unlucky choice of the periodic
orbit by the random-number generator employed to pick symbols. As we
shall see presently, two precautions must be taken for our way
towards ergodic periodic orbits. First, we should fix the
geometric length $L$ rather than the number $n$ of symbols and,
second, wave good-bye to the assumption of uncorrelated symbols.

\subsection{Number of symbols vs. orbit length}

Within an ensemble of orbits of fixed number of symbols $n$ the
orbit length $L$ will fluctuate. Ascribing equal probability to
all allowed symbol sequences and invoking, for large $n$, the
central limit theorem we have the distribution of $L$ as a
Gaussian, with mean and variance both proportional to $n$,

\begin{equation}
g(L|n)   = \frac{1}{\sqrt{2 \pi n \Delta}}\, {\rm e}^{\frac{(L - n
d_n)^2}{ 2 n \Delta}}  , \ \ \ \ \ \ \ \  \int g(L|n) d L = 1 \,;
\label{Gauss}
\end{equation}
$d_n$ and $\Delta$ are the mean length and variance per
symbol. The values of these quantities are system specific; using an ensemble of 100000 periodic
orbits with $n = 100000$ randomly selected symbols each for
the regular octagon we numerically find

\begin{equation}
\label{numval} d_n \approx 2.2568,\quad \Delta \approx 0.6283 \,.
\end{equation}
The concentration of $g(L|n)$ in the vicinity of its maximum
becomes ever more pronounced as $n$ grows. The mean length per
symbol for a long periodic orbit obtained by throwing honest dice
for its symbols will therefore almost certainly be very close to
$d_n$.

Of greater interest is a different ensemble of periodic orbits,
which has the length interval $(L)$ fixed rather than
the number of symbols $n$. In particular, it is the fixed-length
ensemble that one has in mind when speaking about the overwhelming
prevalence of ergodic orbits. To find the number of orbits $N(L) d
L$ in the length interval $(L, L + d L)$ we need the number of
different periodic orbits with $n$ symbols,

\begin{equation} \nu (n) = \frac{1}{n} p_{\rm eff}^n \,,
\end{equation}
where the effective number of symbols $p_{\rm eff} = 6.98$
deviates only slightly from the naive estimate $8-1=7$  obtained
by excluding $k {\bar k}$ patterns; the slight deviation is due to
the group identity (\ref{ide}); the factor $1/n$ reflects the
equivalence of cyclically permuted symbol sequences; a detailed
discussion of $p_{\rm eff}$  will be presented in Appendix
\ref{peffA}. We proceed to $N(L)$ by summing over all possibilities
for the length to lie in the interval $(L, L + d L)$ as
$N(L)=\sum_ng(L|n)\nu(n)$ or, approximating the sum by an
integral,

\begin{equation}
N(L) = \int d n \  g(L|n)\nu(n) = \int d n \frac{1}{\sqrt{2 \pi
n^3 \Delta}}\,{\rm e}^{n \ln p_{\rm eff} - \frac{(L - n d_n)^2}{ 2 n
\Delta}} \label{fixedL}\,.
\end{equation}
In the interesting case of large $L$ the integral can be evaluated
using the saddle-point approximation. The saddle of the
integrand lies at

\begin{equation}
\label{maxq} n_{\rm max} = \frac{L}{\sqrt{d_n^2 - 2 \Delta \ln
p_{\rm eff}}}\,.
\end{equation}
\begin{figure}
\begin{center}
\leavevmode \epsfxsize=0.4 \textwidth \epsffile{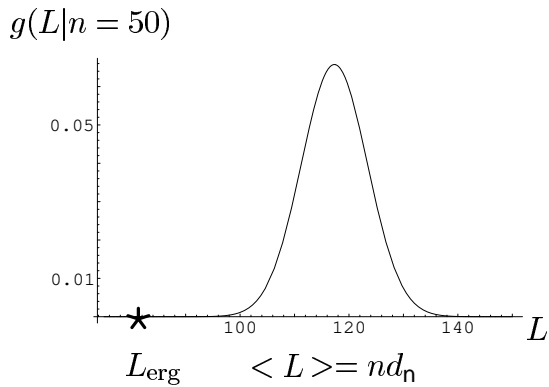}
\end{center}
\caption{For fixed number $n$ of symbols, the 
distribution of orbit lengths is Gaussian (\ref{Gauss})
with maximum at $<L>= n d_n$. However, ergodic orbits are
those at $L_{\rm erg} = n d_L$, in ultra-left wing of
Gaussian.} \label{saddle}
\end{figure}
The orbit density is thus found to obey  the familiar exponential
proliferation law
\begin{equation}
N(L) =\frac{1}{L} {\rm e}^{\eta L}\,, \label{Huber}
\end{equation}
with the growth rate

\begin{equation}
\label{etaeq1} \eta = \frac{d_n - \sqrt{d_n^2 - 2 \Delta \ln
p_{\rm eff}}}{\Delta} \,.
\end{equation}
The latter rate must coincide with the topological entropy. For
billiards on surfaces of constant negative curvature that entropy
equals  unity and (\ref{Huber}) becomes Huber's law \cite{austei0}. With our
approximate numerical values (\ref{numval}) inserted, our
expression  (\ref{etaeq1}) yields $\eta \approx 1.00026$, in nice
agreement with the value $\eta = 1$ required by Huber's law.

The orbits described by the proliferation law (\ref{Huber}) are known
to be ergodic in their overwhelming majority. Since we have just
seen this law to arise mostly from contributions of orbits with
the number of symbols close to $n_{\rm max}$, it is clear that the
ergodic orbits must have the mean length per symbol given by

\begin{equation}
\label{dmax} d_L=\frac{L}{n_{\rm max}} = \sqrt{d_n^2 - 2 \Delta
\ln p_{\rm eff}} \,.
\end{equation}
For the regular octagon, with its values for $d_n, \Delta$ given
in (\ref{numval}) we have $d_L= 1.6283$. Note that this latter value
strongly deviates from the mean length per symbol $d_n= 2.2568$
obtained from randomly chosen sequences of $n$ symbols. Therefore,
orbits generated using random sequences of $n$ symbols have an
exceedingly small probability to be ergodic. To see this we must
realize that the orbits constituting the maximum of the integrand
in the distribution at fixed length (\ref{fixedL}) (known to be
ergodic) are not those forming the maximum of the Gaussian
distribution (\ref{Gauss}) at fixed number of symbols, but rather
belong to its ultra-left wing; the difference of the locations of
the maxima of the two distributions in question is of order $L$,
while the widths are only of order $\sqrt{L}$.

The enormous difference between the two ensembles is due to the
exponential growth of the total number of orbits with $n$ symbols,
$\nu(n)$. The orbits with the length per symbol close to $d_L$ are
in extremely small proportion to all orbits with the same number
of symbols. However, they are dominant among the orbits with a
given length since they have many more symbols.

The fact that the maximum of the integrand in (\ref{fixedL}) is
formed by ergodic orbits can be regarded as due to a theorem of
Bowen's \cite{Knieper}. The number of orbits within the ergodic
subset (characterized by $d_L$) grows in the limit $L \to \infty$
as fast as the total number of orbits.

\subsection{Symbol correlations in ergodic orbits}

Correlations must be accounted for in the symbol sequence if
ergodic orbits are to be constructed. We propose to establish such
correlations with the help of geometric considerations.
The number $d N_{\rm \phi}$ of times an ergodic orbit
intersects a line element of infinitesimal length $dl$, with the crossing  angle in the
interval $(\phi,\phi+d\phi)$, obeys
\begin{equation}
\label{dNphi}
d N_{\phi} \propto  \sin \phi \ dl d\phi\,.
\end{equation}
Indeed, $dN_{\phi}$ and the distribution of
self-crossing angles $P(\epsilon |L)$ must have the same sinusoidal
angle dependence; for self-crossings the line element $dl$ is a
stretch of the periodic orbit itself rather than an element of
some fixed line in the billiard. The behavior in question is also similar to
the "cosine law" of optics for the angular distribution of the
intensity of light emitted diffusely by some surface element.

\begin{figure}
\begin{center}
\leavevmode \epsfxsize=0.25 \textwidth \epsffile{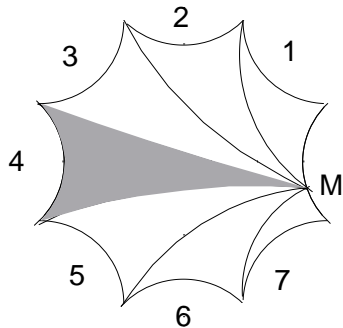}
\end{center}
\caption{Geodesics connecting point $M$ at side
$0$ with corners divide octagon into seven sectors. Orbit starting from
$M$ with momentum direction within $k$-th sector will
arrive at side $k$. Integrating over all such directions and all starting points $M$ on side $0$
gives transition probability $P_{4 \to k}$.}
\label{uebergang}
\end{figure}

We now apply the rule (\ref{dNphi}) to an element of one side of the
regular octagon, say number $0$. The segment of
an ergodic orbit launched from that element
succeeds the orbit segment with the symbol $4$.
In search is now the probability for the orbit segment launched
from side 0 to land on side $k$, i.e. the
probability $P_{4 \to k}$ that a certain symbol $k \neq 0 $
follows the symbol $4$ in the symbolic code. An orbit
segment is uniquely determined by the position of its initial
point on, and the crossing angle  with, the ``launching side''
$0$. We thus find the transition probability  by first integrating (\ref{dNphi}) over the angle
range spanned by side $k$ with respect to the launching element, and subsequently integrating
along side $0$. Fig.~\ref{uebergang} explains the pertinent geometry.

For an irregular octagon the procedure just explained must be
applied to all launching sides $k_{\mbox{in}}$ to  get the
probabilities for the symbol $ k_{\mbox{in}}$ to be
followed by $k_{\mbox{fin}}$. In the regular octagon the
transition probabilities depend only on the absolute value of
$k_{\mbox{in}} -k_{\mbox{fin}}$ which can be 1,2 or 3; they are
to be compared with the value $1/7 = 0.143 \ldots$  which would
correspond to the random symbol sequence without correlation.

Alternatively, symbol correlations can be read off non-periodic trajectories with random initial conditions. Both approaches lead to good agreement for the single-step
transition probabilities $P_{i\to j}$, as documented in the following table,
\begin{eqnarray}\label{transrates}
\begin{tabular}{l ||c |c | c | c |}   &
$P_{4 \rightarrow 1}
$ & $P_{4 \rightarrow 2}
$ & $P_{4 \rightarrow 3}
$  & $ P_{4\rightarrow 4}$  \\ \hline theory & $0.6201$  &
$0.2089$  &  $0.1200$   & $0.0510$
\\ nonper. traj. & $0.6193$  &  $0.2091$  & $0.1205$   &  $0.0512$
\end{tabular}\,.
\end{eqnarray}

Symbol sequences respecting the single-step correlations just explained yield
periodic orbits exhibiting ergodicity to an
impressive degree of approximation. In particular, the average
length per symbol of periodic orbits in the regular octagon
decreases from 2.257 to 1.694, i.e. a value rather close to the
ergodic limit $d_L=1.628$.

To further faithfulness to ergodicity we must account for higher correlations
in the symbol sequence. The next
step involves two-step transition probabilities $P_{i\to j\to k}$, i.e. the conditional
probabilities for a certain symbol to follow any given two-symbol
sequence. For the regular octagon two-step correlations
are embodied in a set of 16 numbers which can be calculated upon
suitably refining the  geometric reasoning applied above to find the four
transition probabilities comprising the single-step correlations. By imposing
two-step correlations on the symbol sequence  we brought the
average length per symbol for periodic orbits down to 1.643; the
coverage of configuration space thus obtained is almost uniform, as
required by ergodicity; see Fig.~\ref{cake}(b). It is worth noting that $P_{i\to j \to k}$ differ from the product of the single-step transition probabilities
$P_{i\to j}P_{j\to k}$ by up to 30\%.

An interesting alternative way to account for all correlations of importance is
to read off symbol sequences of the desired length from a
non-periodic trajectory. The symbolic word  thus obtained  can  be
used for generating an almost perfectly ergodic periodic orbit
which  will almost everywhere  coincide, to a high precision, with
the corresponding section of the non-periodic trajectory. Only a
dozen or so segments of the periodic orbit, namely those
corresponding to the beginning and the end of the word, will
deviate from the non-periodic progenitor. The ergodicity of the orbit thus produced will be close to perfection, as witnessed by the density distribution in Fig.~(\ref{cake})(c). Moreover, the two trajectories
under discussion will have very nearly the same
length per symbol; in fact, considering $10^5$ orbits with $10^5$ symbols each,
generated from non-periodic trajectories, we got the average length per symbol as 1.614. The
latter value differs slightly from the prediction
$d_L= 1.628$ based on (\ref{dmax}) and the
numerical data (\ref{numval}) on the ensemble of the orbits with
random symbol sequences; we have not bothered to look  for an explanation of that
(minute) discrepancy.

\begin{figure}
\begin{center}
\leavevmode \epsfxsize=0.55 \textwidth \epsffile{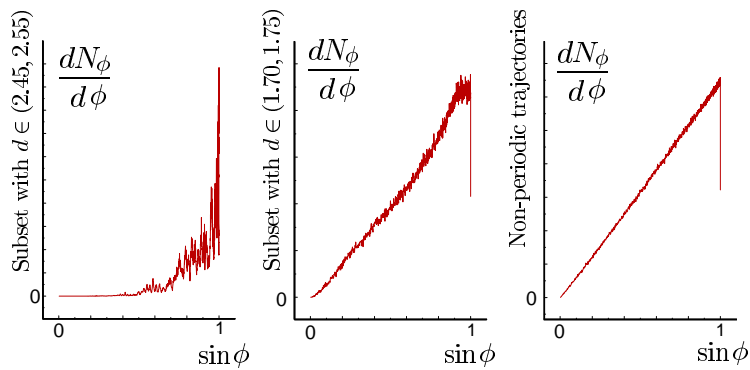}
\end{center}
\caption{Number $d N_{\rm \phi} / d \phi$ of segments 
landing near some point of side $0$ with angle $\phi$ vs $\phi$, for two different values of mean length  $d$ per symbol and for non-periodic trajectories. Fidelity to ergodic sine law 
(\protect\ref{dNphi}) increases as $d$ approaches $d_L$ of (\protect\ref{dmax}).}
\label{eit}
\end{figure}

To finally illustrate the suitability of the mean length per symbol as an indicator
of the fidelity of periodic orbits to ergodicity, Fig.~\ref{eit} depicts the distribution
$dN_{\phi}/d\phi$ of the angle $\phi$ between landing side and arriving periodic orbit, for
groups of orbits of different mean lengths $d$ per symbol. Such groups were constructed from
an ensemble of periodic orbits pertaining to a large fixed number of random uncorrelated
symbols. The distribution of lengths $L$ within such an ensemble is the Gaussian depicted in
Fig.~\ref{saddle}. The approach to ergodicity with $d\to d_L$ is clearly visible in
Fig.~\ref{eit}

\section{Crossing angles, avoided-crossing widths, and loop lengths}\label{joint}

We can now turn to the statistics of self-crossings and associated avoided crossings which is believed to be an important classical ingredient in the explanation of universality of fluctuations in quantum energy spectra of chaotic dynamics \cite{gedisieb,siebrich,qugra,BHH}.

\subsection{Crossings}\label{cross}

\begin{figure}
\begin{center}
\leavevmode \epsfxsize=0.1 \textwidth \epsffile{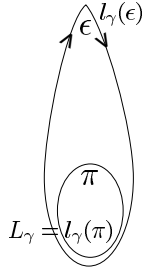}
\end{center}
\caption{Deformation of loop to periodic orbit.} \label{deformation}
\end{figure}

Each crossing divides a periodic orbit of length $L$ into two
loops whose lengths $l_1, l_2$ add up to $L$. In what follows, we
shall focus on the shorter one and its length $l= {\rm min}
(l_1,l_2)$. For the statistical considerations to follow we employ
the number of loops $p(\epsilon, l | L) d\epsilon  \ d l $ with
lengths in the interval $(l,l + d l)$ and crossing angles in the
interval $(\epsilon, \epsilon + d \epsilon)$ in ergodic orbits of
length $L$.   The density of loop
lengths and crossing angles, $p(\epsilon,l|L)$, be calculated
analytically for the Hadamard-Gutzwiller model due to the fact
that each loop can be continuously deformed to a loop with crossing
angle $\pi$, that is, to a periodic orbit $\gamma$, as indicated
in Fig.~\ref{deformation}. Due to the equality of the Lyapunov exponents
for all trajectories in the Hadamard-Gutzwiller model the loop
length $l_\gamma$ is uniquely determined by the crossing angle
$\epsilon$ and the length $L_\gamma$ of the periodic orbit
$\gamma$ as \cite{gedisieb}
\begin{equation}\label{1.2}
\cosh \frac{l(\pi)}{2} \equiv \cosh\frac{L_\gamma}{2} = \cosh\frac{l(\epsilon)}{2}
\sin\frac{\epsilon}{2}\,;
\end{equation}
for $L_{\gamma}$ large that relation simplifies to

\begin{equation}
\label{smeps} l_\gamma(\epsilon)-L_\gamma \approx  2\ln
\frac{1}{\sin \frac{\epsilon}{2}}\,.
\end{equation}
We immediately infer that a loop is longer than its periodic-orbit
deformation $\gamma$ by an amount independent of $\gamma$ and the
longer the smaller the crossing angle; even for the shortest
orbits with the length $L_0=3.06$ (the value refers to the regular
octagon) the error of the simplified relation (\ref{smeps}) relative
to  (\ref{1.2}) is less than 2\%.

Using the basic relation (\ref{1.2}), Sieber expresses $p(\epsilon, l | L)$ as a sum over periodic orbits \cite{gedisieb},

\begin{equation}
\label{dds} p(\epsilon, l | L) = \frac{L}{2\pi A} \sum_{\gamma}
\frac{ \ L_{\gamma} \ \delta(l - l_{\gamma}(\epsilon)) \sin
\epsilon} {r_\gamma\sqrt{ (-1+\cosh L_\gamma)(-1+\cosh L_\gamma+
2 \cos^2\frac{\epsilon}{2}) }}\;;
\end{equation}
here $r_\gamma$ is a repetition number equaling 1 if $\gamma$ is a primitive periodic orbit,
and $A = 4 \pi$ is the area of the octagon. Due to $l_{\gamma}(\epsilon)>L_{\gamma}$
only orbits with $L_\gamma< l< L/2$ contribute to the sum.

As $l$ grows, more and more periodic orbits contribute and the sum
will eventually be dominated by long orbits for which $2\cosh
L_\gamma\approx \exp L_\gamma \gg 1$ so that the denominator may
be replaced  by $\exp L_\gamma$.  Taking into account (\ref{smeps})
and applying the sum rule of Hannay and Ozorio de Almeida \cite {HOdA} we
obtain the ergodic result,

\begin{equation}
\label{ergo} p(\epsilon, l | L) \approx \frac{L}{\pi A} \sin
\epsilon, \quad l\gg l_0 (\epsilon)\;,
\end{equation}
where $\cosh(l_0(\epsilon/2))= \cosh (L_0/2)/\sin(\epsilon/2)$
gives  the length of the shortest possible loop created by
deformation of the shortest periodic orbit of the system with
length $L_0$, again according to the basic relation (\ref{1.2}).

\begin{figure}
\begin{center}
\leavevmode \epsfxsize=1.00 \textwidth \epsffile{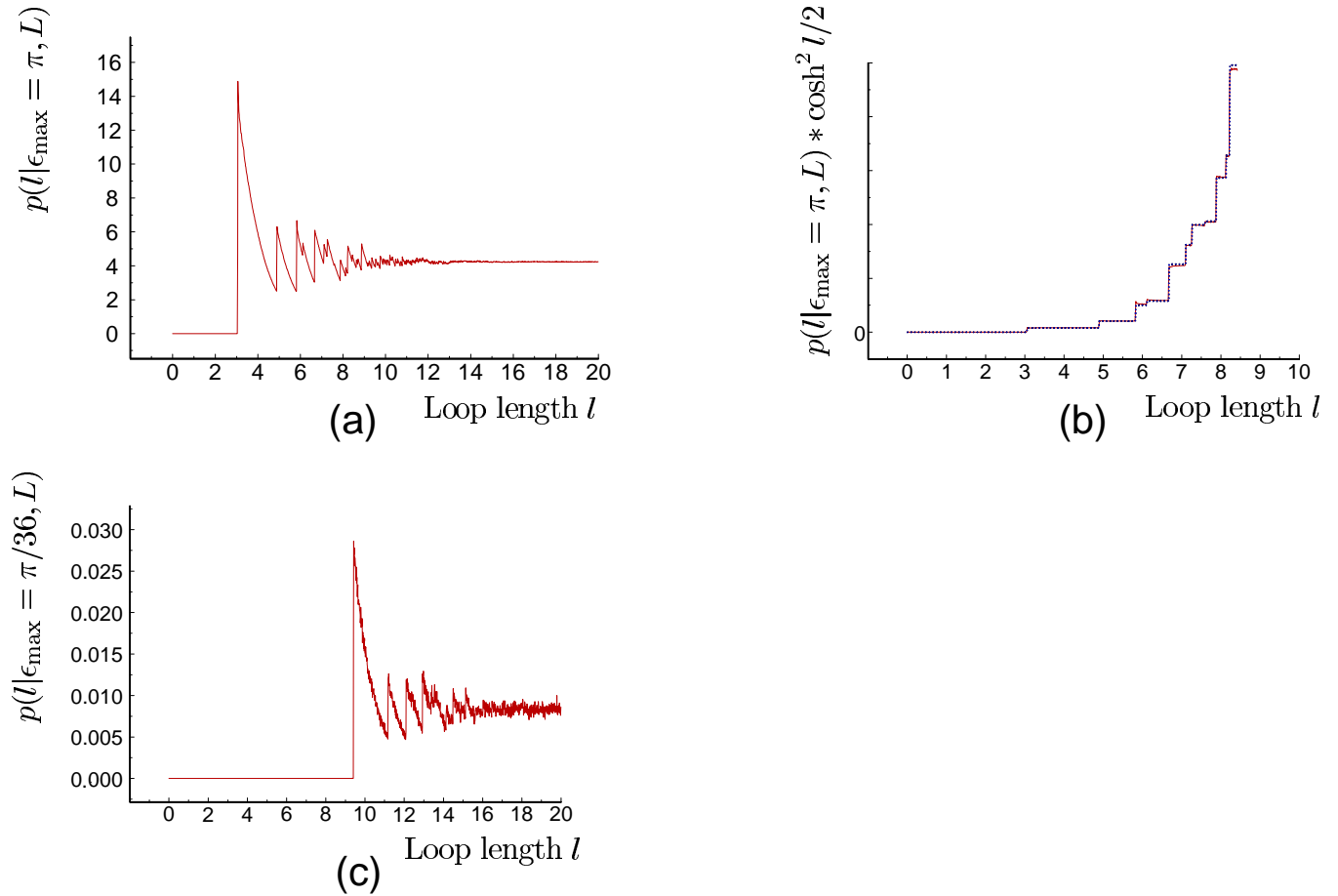}
\end{center}
\caption{Joint density of loop lengths and crossing angles; analytic result
(\ref{lolgh}) for $\epsilon_{\rm max} = \pi$ and $\epsilon_{\rm max}
= \pi/36$ indistinguishable from numerical finding. Angle-dependent gap visible in
(a) and (c) is origin of logarithmic correction to 
ergodic behavior of self-crossing distribution.} \label{loop}
\end{figure}

Two reduced distributions can be obtained from the general
expression (\ref{dds}). Integrating over the angle from zero to some
 $\epsilon_{\rm max}$, we obtain the length distribution
$p(l |\epsilon_{\rm max},L)$ for loops with crossing angles
smaller than $\epsilon_{\rm max}$ in the periodic orbit of length
$L$. The resulting expression is astonishing simple,

\begin{equation}
\label{lolgh} p(l |\epsilon_{\rm max}, L) = \int_0^{\epsilon_{\rm
max}} p(\epsilon, l | L) d \epsilon =
\frac{L}{2\pi A\cosh^2 \frac{l}{2}} \ \sum_{ l_{\gamma}(\epsilon_{\rm
max}) \leq l} \ \frac{L_{\gamma}}{r_\gamma} {\rm coth}
\frac{L_{\gamma}}{2}
\,.
\end{equation}
The only $\epsilon_{\rm max}$-dependence lies in the summation condition
$l_{\gamma}(\epsilon_{\rm max}) \leq l$ and indicates that only those periodic orbits
$\gamma$ contribute to the sum which can be deformed to a loop of
length $l_{\gamma}(\epsilon_{\rm max})$ smaller than the given
length $l$.  Therefore, we obtain a staircase function for the
quantity $p(l |\epsilon_{\rm max}, L){\cosh^2 l/2}$.  In Fig \ref{loop} (a), (c),
we display $p(l |\epsilon_{\rm max}, L)$ for $\epsilon_{\rm max} = \pi$ and
$\epsilon_{\rm max} = \pi/36$, whereas (b) displays the staircase
obtained from (a) after multiplication with $\cosh^2 l/2$; these
plots reveal perfect agreement between the numerical results
obtained from all loops in an ensemble of $10^6$ ergodic periodic
orbits with the  prediction of  the periodic-orbit sum
(\ref{lolgh}).

Note that a gap arises in the length distribution $p(l
|\epsilon_{\rm max}, L)$ due to the existence of a shortest
periodic orbit and the related shortest possible loop of angle
$\epsilon$ and length $l_0(\epsilon)$. The gap is  given by $0\leq
l\leq l_0 (\epsilon_{\rm max})$. In particular, in
Fig.~\ref{loop}(a), $l_0(\epsilon_{\rm max}= \pi) = L_0 = 3.06$ is
just the length of the smallest periodic orbit of the system,
whereas in Fig.~\ref{loop} (c), $l_0 (\epsilon_{\rm max}=
\frac{\pi}{36}) = 9.41$ is the length of the shortest possible
loop with angle $\epsilon_{\rm max}= \pi/36$. Generally, in accord
with the relation (\ref{smeps}) between the lengths
$l_{\gamma}(\epsilon)$ and $L_{\gamma}$, the  peaks related to all
periodic orbits experience practically the same shift to the
right, $\sim 2\ln(2/\epsilon)$; therefore, when $\epsilon_{\rm
max}$ is decreased  the length distribution appears to shift to
the right rather rigidly.

When (\ref{dds}) is integrated over all loop lengths $l$, we obtain
the density of crossing angles $P(\epsilon | L)$ in ergodic
periodic orbits of length $L$ as

\begin{eqnarray}
P(\epsilon | L)  = \int_0^{L/2} p(\epsilon, l | L) d l = \frac{L
\sin \epsilon}{\pi A}\left(\frac{L}{2} - l_{\rm eff}
(\epsilon)\right)
 \,.
\label{lout}
\end{eqnarray}
The  term in the right hand side proportional to $L^2$ could be
obtained immediately using the ergodic approximation (\ref{ergo}).
The actual number of crossings is smaller due to the already
mentioned gap in the loop length distribution, such that the
effective integration interval in (\ref{lout}) is shorter than $L/2$.
This is taken into account by the introduction of the  effective
minimal loop length $l_{\rm eff} (\epsilon)$; that latter quantity
is close to, but  somewhat smaller than, the  gap visible in Fig
\ref{loop}(a), (c), i.e. the length $l_0(\epsilon)$ of the loops
related to the shortest periodic orbit. The discrepancy arises
from the shortest periodic orbits whose contribution is greatly in
excess of  the  ergodic prediction.

A periodic-orbit sum for $l_{\rm eff} (\epsilon)$ is obtained by
substituting the  loop length density (\ref{dds}) for the integrand
in (\ref{lout}) and comparing the left and right hand sides.  Recalling
that all loop lengths exceed the lengths of their associated periodic
orbits by the same quantity (see (\ref{smeps})) and employing the sum
rule, we can make the summation interval independent of $\epsilon$,
 \begin{eqnarray}
l_{\rm eff}(\epsilon)&=&2\ln \frac {1}{\sin \frac \epsilon
2}+\phi(\epsilon);\nonumber\\ \phi(\epsilon)&\equiv& \frac L 2 -
 \sum_{L_\gamma<\frac L 2}
\frac{  L_{\gamma} } {r_\gamma\sqrt{ (-2+2\cosh
L_\gamma)(-2+2\cosh L_\gamma+ 4 \cos^2 \epsilon/2) }}\;.
\label{lmin2}
 \end{eqnarray}

For long orbits (large $L$) the last expression is $L$-independent. Indeed, the
increment of $\phi(\epsilon)$, when $L$ is replaced by $L+\Delta L$, is given by
\begin{equation}
\Delta\phi\approx\frac{\Delta L}{2}-\sum_{\frac{L}{2}<L_{\gamma}<\frac{L+\Delta L}{2}}L_\gamma {\rm e}^{-L_\gamma}=0,
\end{equation}
which follows from the proliferation law for periodic orbits.

The $\epsilon$ dependence of $\phi(\epsilon)$ comes from the
shortest and thus non-generic orbits; it is so weak that for most
purposes we can set $\phi(\epsilon)\approx \phi(0)$ and work with
\begin{equation}\label{leffprox}
l_{\rm eff}(\epsilon)\approx 2 \ln \frac{c}{2\sin\frac \epsilon
2},\quad c= 2\mbox{e}^{\frac {\phi(0)}{2}}\approx 3.16\,;
\end{equation}
the constant $c$ was obtained by  choosing  $L/2=12$ and
correspondingly allowing  for periodic orbits with lengths
$L_\gamma\le 12$ in $\phi(0)$.

\subsection{Avoided Crossings}\label{avcross}

Inasmuch as every periodic orbit with a small-angle
self-intersections has a partner orbit which is almost identical
save for avoiding the said crossing, one would expect the
statistics of crossings and avoided crossings to be the same, at
least in the limit of small angles. In the present subsection we
shall confirm that expectation, mainly by showing that there is a
formal analytic-continuation relationship between crossings and
avoided crossings. However, it is important to notice that two
different types of avoided crossings arise which have different
properties.

\begin{figure}
\begin{center}
\leavevmode \epsfxsize=0.55 \textwidth \epsffile{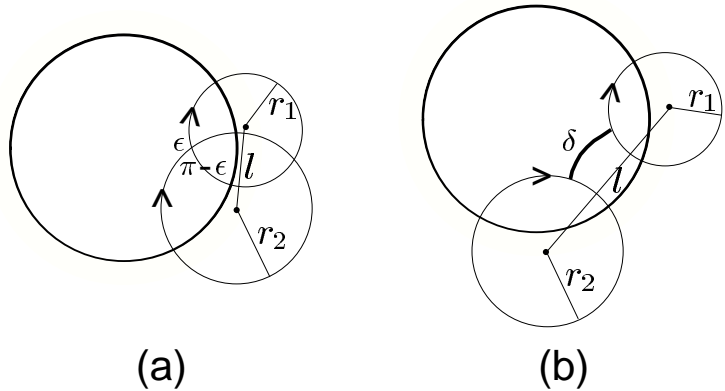}
\end{center}
\caption{Geodesics in unit disc may cross once (a), or form avoided crossing. Part (b) depicts antiparallel  avoided crossing;
reversion of motion in one geodesic would interchange angles as $\epsilon\leftrightarrow\pi-\epsilon$ in (a) and form  parallel avoided crossing in (b).} \label{geom}
\end{figure}

We generalize the considerations of the preceding subsection by
imagining the division of a periodic orbit into two ``loops'' by
an avoided crossing. As shown in Fig.~\ref{geom}, two geodesics,
the intra-octagon segments of which belong to a periodic orbit,
may either cross inside the unit disc once or not at all. As before, we let $\epsilon$ be the
angle complementary to the angle between the velocity vectors of
the two geodesics at an intersection as defined in Fig.~\ref{geom}
(a). When the direction of velocity in one of the two geodesics is
reversed, the role of the angles $(\epsilon, \pi-\epsilon)$ is
interchanged. Elementary
Euclidian geometry provides the relation between the radii $r_1,
r_2$ of the two geodesic circles, the distance $l$ between its
midpoints and the angle $\epsilon$ as

\begin{eqnarray}
\sin^2 \frac{\epsilon}{2} = \frac{1}{2} \pm \frac{r_1^2 + r_2^2 -
l^2} {4 r_1 r_2}\,, \label{huhu}
\end{eqnarray}
where all quantities are measured in Euclidian units. The two
solutions with $\pm$ correspond to the two complementary angles
$(\epsilon, \pi-\epsilon)$.

In case the two segments do not cross there exists a minimal
distance $\delta$ between them. That distance is measured using
the metric (\ref{metric}), along the geodesic
orthogonal to the said segments.  The two
intersection points of this geodesic with the segments of the
orbit divide the orbit into two loops, much the same as in the
case of a crossing.
 Surprisingly, the distance $\delta$ is given
by analytic continuation ($\epsilon\to{\rm i}\delta$) of equation
(\ref{huhu}),

\begin{eqnarray}
\sinh^2 \frac{\delta}{2} = - \frac{1}{2} + \Big| \frac{r_1^2 +
r_2^2 - l^2} {4 r_1 r_2} \Big| \,, \label{anal}
\end{eqnarray}
where, in contrast to $\delta$, the radii $r_1,r_2$ and the
distance $l$ are still meant in the Euclidian sense, as in
(\ref{huhu}). We do not want to pause and give the straightforward
but dull proof of (\ref{anal}) here but shall actually recover both
(\ref{huhu}) and its analytic continuation (\ref{anal}) through symbolic
dynamics in Sect.~\ref{symbdyn}.

The geodesic along which the width of an avoided crossing is measured may
consist of several disjoint arcs within the fundamental domain; we
shall come back to this complication in Sect.~\ref{partnership}.

In what follows, it will be important to distinguish two different
types of avoided crossings: We call ``antiparallel'' the avoided
crossings with antiparallel velocities at the point of minimal distance as in Fig.~\ref{geom} (b) while
avoided crossings with parallel velocities are called "parallel". We shall be
interested in the respective distributions $p^a(\delta, l | L)$ and
$p^p(\delta, l |L)$ of the
loop length $l$ and the minimal distance $\delta$,  both meant for loops inside
ergodic orbits of total length $L$; clearly, these densities
correspond to the distribution of loop lengths and crossing angles considered in the
preceding subsection.

\begin{figure}
\begin{center}
\leavevmode \epsfxsize=1.0 \textwidth \epsffile{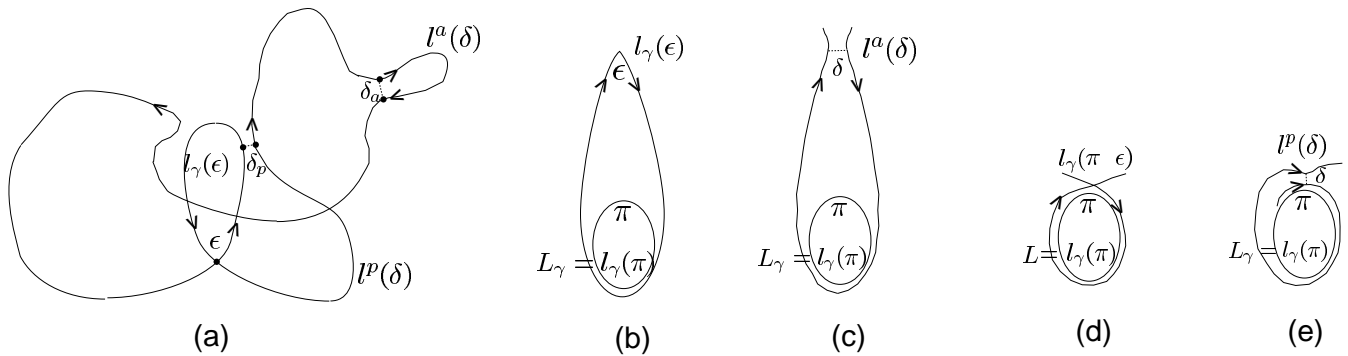}
\end{center}
\caption{Crossings, parallel and antiparallel avoided crossings.
For avoided crossings orbit is divided into loops
by geodesic whose length $\delta$ is width of 
avoided crossing. For narrow antiparallel (c) and
parallel (e) avoided crossings, loop length is almost same as
for crossing with angle $\epsilon = \delta\ll 1$ (b) and $\pi -
\delta$ (d).}
\label{ally}
\end{figure}

When determining the density $p(\epsilon,l | L)$, we exploited the idea that every loop with length $l$ and
crossing angle $\epsilon$ in a given orbit of length $L$ is
related to one periodic orbit $\gamma$, of which it can be viewed
as a deformation; it is that orbit $\gamma$ which fulfills $\cosh
(L_{\gamma}/2) = \cosh (l/2) \sin (\epsilon/2)$. In this sense, we
have the association $(l, \epsilon) \leftrightarrow \gamma$.

Similarly, each loop defined by an antiparallel avoided crossing
can be viewed as a deformation of one particular periodic orbit
$\gamma$ with length $L_{\gamma}$, and the same is true for
parallel avoided crossings. The  analytic expressions for the
relationships $\gamma \leftrightarrow ({l^a}, \delta)$ and
$\gamma' \leftrightarrow ({l^p}, \delta)$ appear as analytic continuations
of (\ref{1.2}) , in analogy to the above
continuation of the simple trigonometric identity (\ref{huhu}) for
the crossing angle to (\ref{anal}) for the closest-approach distance.
The precise meaning of this analytic continuation will be
explained in the following section, see equation (\ref{myfo}). Here,
we just mention that antiparallel avoided crossings correspond to
solutions of Eq. (\ref{1.2}) with $\epsilon$ replaced by $i \delta$,
and parallel avoided crossings to solutions with $\epsilon$
replaced by $\pi + i \delta$. The  lengths
${l^a}_{\gamma}(\delta)$ and ${l^p}_{\gamma}(\delta)$ of loops
generated by avoided crossings as defined in Fig.~\ref{ally}(a)
can be related to the lengths of periodic orbits $\gamma$ as
\begin{eqnarray}
{l^a}_{\gamma}(\delta) &=& 2 \ {\rm arcosh}
\Big(\frac{\cosh{\frac{L_{\gamma}}{2}}} {\sinh{\frac{\delta}{2}}}\Big)
\quad {\rm antiparallel}\,,
\label{anti}\\
{l^p}_{\gamma'}(\delta) &=& 2 \ {\rm arcosh}
\Big(\frac{\cosh{\frac{L_{\gamma'}}{2}}} {\cosh{\frac{\delta}{2}}}\Big)
\quad {\rm parallel}\,.
\label{para}
\end{eqnarray}
Of course, in general $\gamma$ and
$\gamma'$ are different periodic orbits. For small
closest-approach distances $\delta$, the foregoing expressions for
the loop lengths are simplified to
\begin{eqnarray}
{l^a}_{\gamma}(\delta) - L_{\gamma } &  \approx & \  2 \ln
\frac{2}{\delta} \ \approx   \ \ l_{\gamma}(\epsilon) - L_{\gamma}
+ {\cal O}(\epsilon^2) \ \ \ \ {\rm for} \ \epsilon = \ \delta \
<< \ 1 \nonumber \\ {l^p}_{\gamma'}(\delta) & \approx &
L_{\gamma'} \ \ \ \ \ \ \ \ \ \ \ \ \ \ \ \
 \ \ \ \ \ \ \ \ \ \ \ \ \ \ \ \ \ \ \ \ \ \ \ {\rm for} \ \delta \ <<  \ 1\,.
\label{limit}
\end{eqnarray}
In part (b) of Fig.~\ref{ally}, we schematically depict  loops and
their lengths $l_{\gamma}(\epsilon)$  for small crossing angles
($\epsilon << 1$), similarly in part (c) for
${l^a}_{\gamma}(\delta)$ with narrowly avoided antiparallel
crossings ($\delta \ll 1$) , in part (d) for
$l_{\gamma}(\epsilon)$ with large angle $\epsilon \approx \pi$,
and finally in part (e) for ${l^p}_{\gamma}(\delta)$ with narrowly
avoided parallel crossing ($\delta\ll 1)$.

\begin{figure}
\begin{center}
\leavevmode \epsfxsize=0.75 \textwidth \epsffile{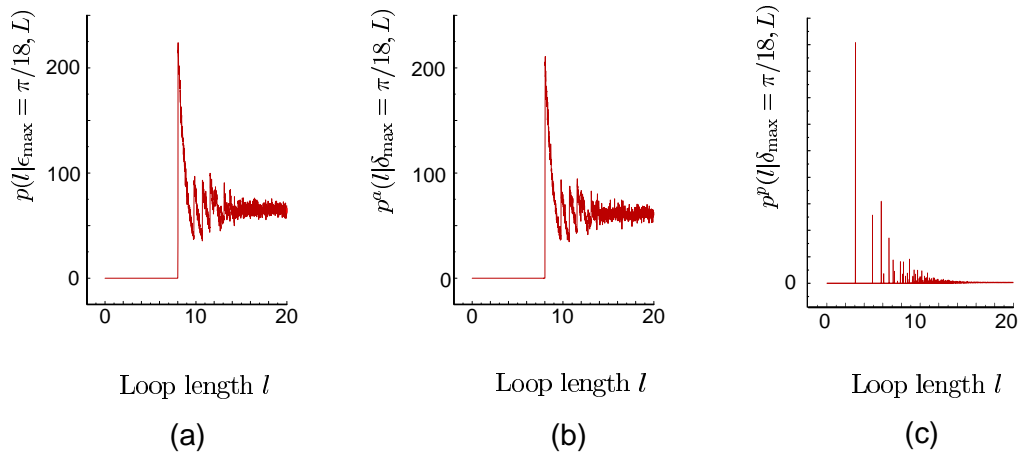}
\end{center}
\caption{ Loop length distributions $p(\epsilon_{\rm max} = \pi/16 | L)$ for self-crossings (a) and $p^a(\delta_{\rm max} = \pi/16 | L)$ for antiparallel
avoided crossings; both practically coincide. Very different looking loop length distribution $p^p(\delta_{\rm max} = \pi/16 |L)$
for parallel avoided crossings in (c) exhibits peaks near periodic orbit lengths since narrow parallel avoided crossings create loops close to
periodic orbits; see Fig. \ref{ally} (e) and (\ref{limit}).}
\label{joop!}
\end{figure}

In analogy to the preceding section, we proceed to the marginal
distributions $p^{p/a}( l |\delta^{\rm max}, L)$  defined as
\begin{equation}
p^{p/a}( l |\delta^{\rm max}, L) = \int_0^{\delta^{\rm max}}
p^{p/a}(\delta, l | L) d \delta \,.
\end{equation}
Most interesting is the limiting case of narrowly avoided
crossings. In Fig.~\ref{joop!} (a) and (b), the loop length
distributions $p( l|\epsilon_{\rm max},L)$ for crossings and $p^a(
l|\delta^{\rm max},L)$ for antiparallel avoided crossings are
displayed with $\epsilon_{\rm max} = \delta^{\rm max} = \pi/18$.
Practically no difference can be observed between these two
distributions, as predicted by $l_{\gamma} \approx l^a_{\gamma}$
in Eq. (\ref{limit}) for small argument; we here encounter the
expected statistical correlation between crossings and avoided
crossings. On the other hand, for parallel avoided  crossings, the
distribution $p^p(l|\delta^{\rm max}, L)$ with $\delta^{\rm max} =
\pi/18$ looks dramatically different: As displayed in (c), spikes
exactly at the position of periodic orbits are observed, as
predicted by ${l^p}_{\gamma}(\delta) \approx L_{\gamma}$ in
(\ref{limit}) in the limit of narrowly avoided parallel crossings.

\newpage

\section{Symbolic dynamics for crossings and avoided crossings}\label{symbdyn}

\subsection{Crossings}
We  here propose to reveal how  self-intersections with small
angles and  narrowly avoided crossings are reflected in
symbolic words. That correspondence will lead
us to some exact results for the Hadamard-Gutzwiller model, as
well as a better understanding of the analytic-continuation
relationship of crossings and avoided crossings. The ideas to be
presented may also become important for generalizations of the
Sieber-Richter theory to other models such as billiards, quantum
graphs and maps.

\begin{figure}
\begin{center}
\leavevmode \epsfxsize=0.45 \textwidth \epsffile{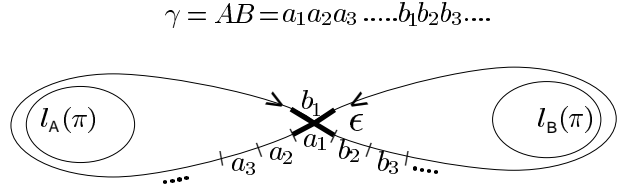}
\end{center}
\caption{Schematic decomposition of self-crossing orbit in segments, one per symbol. First letter
$a_1$ of word $W$ and its part $A$ refers to one of two segments  involved in crossing; similarly, first letter $b_1$ of $B$ refers to other crossing segment.} \label{fuebl}
\end{figure}

In what follows, symbol sequences with brackets $\{ ... \}$ refer
to periodic orbits, whereas sequences without brackets refer to
parts thereof. Consider a crossing with  angle $\epsilon $
dividing a periodic orbit into two loops. The symbolic word of the
orbit will also be divided by the crossing into two parts which
refer to a loop each and will be denoted by $A= a_{1}a_{2}\ldots
a_{n_{a}} $ and $B= b_{1}b_{2}\ldots b_{n_{b}} $ where
$a_{i},b_{k}$ are symbols from 0 to 7; see Fig. \ref{fuebl}. The
word of the whole orbit will be $\{ AB \} =\{a_{1}a_{2}\ldots
a_{n_{a}}b_{1}b_{2}\ldots b_{n_{b}}\}$, with the number of symbols
$n=n_{a}+n_{b}$ and the length equal to the sum of the loop
lengths, $L_{AB}=l_{A}(\epsilon)+l_{B}(\epsilon)$. The word is
assumed pruned (see Section IV).

We recall that the loops $A,B$ can be continuously deformed to periodic
orbits $\{A\},\{B\}$, whose lengths $L_{A},L_{B}\,$
related to the lengths of the loops by (\ref{1.2}), i.e., as
$\cosh\frac{l_{W}(\epsilon )}{2}=\cosh
\frac{L_{W}}{2}/\sin \frac{\epsilon }{2}$ for ${\cal W}={\cal A}
\mbox{ or } {\cal B}$; we shall also have to relate
the lengths of both orbits  to the traces
of the corresponding M\"obius-transformation matrices\footnote{In
contrast to the preceding sections we here found it convenient to
notationally distinguish between symbol sequences $A,B,AB,W$ and
the associated matrices ${\cal A,B,AB,W}$} according to (\ref{1.1}), i.e.
$\cosh\frac{L_{W}}{2}=| \mathrm{Tr}\ \frac{{\mathcal
W}}{2}|$. Here ${\mathcal A,B}$ are the $2\times 2$ matrices obtained by
taking the product of elementary matrices (\ref{elem}), in the order
given by the symbol sequences $A,B$; the foregoing expression for
$L_W$ also holds for non-regular octagons, with the elementary
matrices (\ref{elem}) replaced by the pertinent generators.

Summing up the loop lengths we
get
\begin{equation}
L_{AB}=2\ \mathrm{arcosh}\left| \mathrm{Tr}\ \frac{{\mathcal
A}{\cal B}}{2} \right| =2\ \mathrm{arcosh}\left| \frac{\mathrm{Tr\
{\mathcal A}}/2}{\sin \epsilon /2}\right| +2\
\mathrm{arcosh}\left| \frac{\mathrm{Tr\ {\mathcal B}} /2}{\sin
\epsilon /2}\right|\,,
\end{equation}
and, solving for $\sin^2\epsilon/2$
\begin{equation}
\sin ^{2}\frac{\epsilon }{2}=\frac{(\rm{Tr}\ {\mathcal
AB})(\rm{Tr}\ {\cal A})(\rm{Tr}\ {\cal B})-(\rm{Tr\ {\cal
A}})^{2}- (\rm{Tr\ {\cal B}})^{2}}{(\rm{Tr\ {\cal
AB}})^{2}-4}\equiv F({\cal A},{\cal B})\,, \label{myfo}
\end{equation}
except that the elementary derivation yields the first term in the
numerator of the second member as $|({\rm Tr}\ {\mathcal AB})({\rm
Tr}\ {\cal A})({\rm Tr}\ {\cal B})|$. We have, somewhat
frivolously, dropped the modulus operation since numerical
evidence suggests $(\rm{Tr}\ {\mathcal AB})(\rm{Tr}\ {\cal
A})(\rm{Tr}\ {\cal B})\geq 0$ whenever the loops $A,B$ arise from
a crossing; we regret not having found a proof, all the more so
since only the form conjectured in (\ref{myfo}) allows for analytic
continuation to the case of avoided crossings (see below).
Incidentally, the foregoing symbolic-dynamics expression for the
crossing angle is the previously announced variant of the
elementary-geometry result (\ref{huhu}). For long orbits and loops we
may approximate (\ref{myfo}) as
\begin{equation}
\sin ^{2}\frac{\epsilon }{2}\approx  \frac{\mathrm{Tr}{\mathcal
A}\ \mathrm{Tr} {\mathcal B}}{\mathrm{Tr}{\mathcal AB}}\,.
\label{smaa}
\end{equation}

Inasmuch as a crossing is connected with the loops $A,B$, the
right hand side of (\ref{myfo}) must obey $0<F\,({\mathcal
A},{\mathcal B})<1$. As regards the inverse statement, one must be
a little careful. There can be several divisions of the word $W$ into
parts $A$, $B$ leading to exactly the same value of $F(\mathcal A,
\mathcal B)$. In this situation, the orbit has one and only one crossing with
the angle defined by (\ref{myfo}) (provided $0<F\,({\mathcal A},{\mathcal
B})<1$), and only one of the divisions mentioned actually describes the
loops associated to the crossing.

As an important application of the foregoing relation between
crossing angles and symbolic words we can now find the condition
on the word for  the crossing angle to be small. As obvious from
(\ref{smaa}) we must have $\mathrm{Tr}{\mathcal
A}\mathrm{Tr}{\mathcal B}<<\mathrm{Tr}{\mathcal AB}$. A crude
estimate of these traces  can be obtained by recalling that the
length of an  ergodic orbit with the (pruned!) word $W$ is roughly
proportional to the number $n_W$ of letters in the word,
$L_{W}\sim n_{W}d_L$, where $d_L$ is the average length per symbol
(close to 1.6 in the regular octagon; see section V). Combining
with (\ref{1.1}) we find that the trace of the M\"obius transform
${\mathcal W}$ grows exponentially with $n_W$,
\begin{equation}
\mathrm{Tr} \ {\mathcal W} \sim e^{\,n_W d_L /2}.
\end{equation}
If $A,B$ are by themselves pruned words we have for the
traces  in (\ref{smaa}) $\mathrm{Tr}{\mathcal A} \sim
\mbox{e}^{n_{A}d_L/2},\mathrm{Tr}{\mathcal B} \sim \mbox{e}^{n_{B}d_L/2},\quad
\mathrm{Tr}{\mathcal AB} \sim \mbox{e}^{(n_{A}+n_{B})d_L/2}$ with the
trivial conclusion $\sin ^{2}\frac{\epsilon }{2}\sim 1$; the
crossing angle will not be small then.

Suppose, however, that the matrices ${\mathcal A}$ and ${\mathcal
B}$ possess the structure
\begin{equation}
{\mathcal A}={\mathcal Z}_{1}{\mathcal LZ}_{1}^{-1},\quad
{\mathcal B} ={\mathcal Z}_{2}{\mathcal RZ}_{2}^{-1}
\label{assummoeb}
\end{equation}
with ${\mathcal L},{\mathcal R}$ generic M\"obius transforms
while the ``insertions '' ${\mathcal Z}_{1},{\mathcal Z}_{2}$ are
products of $many$ elementary matrices. Obviously the insertions
disappear from the traces of the matrices ${\mathcal A}$ and
${\mathcal B}$ since $\mathrm{Tr}\ {\mathcal A} \equiv
\mathrm{Tr\,}{\mathcal Z}_{1}{\mathcal LZ}_{1}^{-1}=
\mathrm{Tr}{\mathcal L}$ and similarly for ${\mathcal B}$; they do not cancel in the trace of the
product of ${\mathcal AB}$, however. Denoting ${\mathcal
Z=Z}_{1}^{-1}{\mathcal Z}_{2}$ we have from (\ref{smaa})
\begin{equation}
\sin ^{2}\frac{\epsilon }{2}\approx \frac{\mathrm{Tr}{\mathcal L}\
\mathrm{Tr}{\mathcal R}}{\mathrm{Tr}{\mathcal LZRZ}^{-1\,\ }},
\end{equation}
and can thus approximate the crossing angle as
\begin{equation}
\sin ^{2}\frac{\epsilon }{2} \approx \frac{\epsilon ^{2}}{4}\sim \mbox{e}^{-n_{Z}\,%
d_L},\quad n_{Z}\equiv n_{Z_{1}}+n_{Z_{2}}.
\end{equation}
Very long insertions ${\mathcal Z}_{1},{\mathcal Z}_{2}$ will
therefore yield very small crossing angles.

\begin{figure}
\begin{center}
\leavevmode \epsfxsize=0.45 \textwidth \epsffile{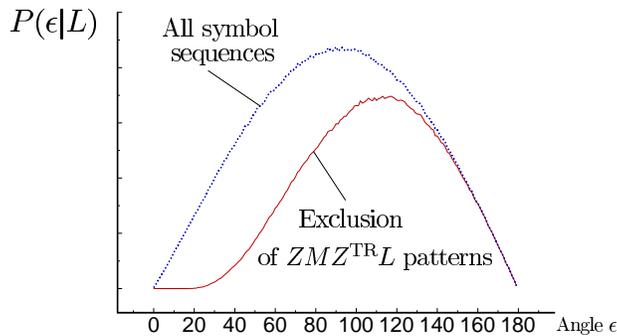}
\end{center}
\caption{Self-crossing distribution $P(\epsilon | L)$ for
ergodic orbits, and for subset of ergodic orbits with patterns $Z M Z^{\rm TR} L$ excluded. $All$
small-angle self-crossings disappear for latter subset.} \label{noangles}
\end{figure}

To understand the physics behind condition (\ref{assummoeb}) we
recall from Section II that inverting a M\"obius matrix
means time reversal for the corresponding code; and that the time
reverse of the code $Z= z_{1}\ldots z_{n} $ is
$Z^{\mathrm{TR}}=\bar{z}_{n}\ldots \bar{z}_{1} $ with
$\bar{z}=(z+4){\rm\mbox{ }mod\mbox{ }}8$.  Condition (\ref{assummoeb}) for the
M\"obius matrices therefore means
\begin{equation}
A=Z_{1}LZ_{1}^{\mathrm{TR}},\quad  B=Z_{2}RZ_{2}^{\mathrm{TR}}
\label{loopstruct}\,.
\end{equation}
The code of the total orbit can be written as
$\{LZRZ^{\rm{TR}}\}$, with $Z=Z_{1}^{\rm{TR}}Z_{2}\,$. The shorter
periodic orbits associated with the loops will have the codes $\{
L \} $ and $\{ R \}$. Since cyclic permutation is allowed in the
code of orbits (not in loops!), the insertions can be brought
together to form the self-canceling sequences
$Z_{1}Z_{1}^{\mathrm{TR}}$ in $\{ A \}$ and
$Z_{2}Z_{2}^{\mathrm{TR}}$ in $\{ B \}$. Using the locality of the
code-to-trajectory connection (see Section III) we may say,
somewhat loosely, that the loop $A$ is obtained by smoothly
joining the periodic orbit $\{ L \}$ with two pieces of
trajectory, one visiting the boundaries of the octagon according
to $Z_{1}$, and the other, encoded by $Z_{1}^{\mathrm{TR}}$,
running through the same sequence of boundaries in the opposite
direction. Such is precisely the behavior of a
loop with a small crossing angle: The sections in the vicinity of the
crossing nearly coincide up to the sense of traversal.

The loop structure (\ref{loopstruct}) seems in fact
necessary to produce a small crossing angle. Indeed, the beginning
and ending parts of a loop being ``nearly parallel'' near a
small-angle crossing the corresponding symbol sequences cannot
help being locally identical, apart from time reversal.
Impressive numerical evidence for the equivalence is provided by
Fig.~\ref{noangles}: Depicted are the densities $P(\epsilon |L)$
of crossing angles for a large generic set of symbol sequences and
the subset purged of sequences of the indicated type; in the
latter case no crossings with small angles are present.

\subsection{Avoided crossings}

The symbolic dynamics of the avoided crossings is investigated in
much the same way. One has to replace the loop length
(\ref{1.2}) by the corresponding results  (\ref{anti}) and (\ref{para}).
That replacement suggests an interpretation as the
analytical continuation $\varepsilon \rightarrow i\delta $ and
$\varepsilon \rightarrow \pi +i\delta $ for
antiparallel and parallel avoided crossings, respectively. The
width of an avoided crossing dividing the orbit
into the loops $A,B$ becomes
\begin{eqnarray}
\sinh ^{2}\frac{\delta }{2\,} &=&-F({\mathcal A,B)}\approx
-\frac{\mathrm{Tr}{\mathcal A}\ \mathrm{Tr} {\mathcal
B}}{\mathrm{Tr}{\mathcal AB}}  \ \ \ \ \mbox{ antiparallel ,} \label{aac}
\\ \cosh ^{2}\frac{\delta }{2\,}
&=&+F({\mathcal A,B)} \approx  +\frac{\mathrm{Tr}{\mathcal A}\
 \mathrm{Tr}{\mathcal B}}{\mathrm{Tr}{\mathcal AB}} \ \ \ \
\mbox{ parallel} \label{pac}\,,
\end{eqnarray}
where the approximate equalities refer to long orbits and loops,
as in (\ref{smaa}). It is obviously necessary that $F({\mathcal
A,B})<0$ \ ($F({\mathcal A,B})>1)$ \ for the existence of an
antiparallel (resp. parallel) avoided crossing. The width of the
antiparallel avoided crossing will be small under the same
assumptions (\ref{loopstruct}) about the loops as in the case of
small-angle crossings; the code of the whole orbit  in both
cases has the structure $LZRZ^{\mathrm{TR}}$  such that the
distinction can be made only after calculating and thus determining the sign of
$F({\mathcal A,B)}$.

The Sieber-Richter theory is based on the fact that each orbit
with a small crossing angle $\epsilon $ has a partner  with almost
the same length which avoids that crossing
by the width $\delta \approx\epsilon$
(see the next subsection for precise relations between the loop
lengths, $\epsilon$, and $\delta$); the partner has practically
the same loops but one of them is traversed in the opposite sense
relative to the original orbit. We can now translate these
assertions into the language of symbolic dynamics. As we have just
seen, a small-angle crossing divides the orbit into loops $A,B$
with the structure (\ref{loopstruct}); consequently the partner
must consist of the loops $A,B^{\mathrm{TR}}$.
Accounting for long insertions near the crossing as after (\ref{loopstruct})
we can state that each Sieber-Richter pair of orbits with small
$\epsilon$ and $\delta$ must have codes $LZRZ^{\mathrm{TR}}$ and
$LZR^{\mathrm{TR}}Z^{\mathrm{TR}}$.  The
crossing angle in one of the orbits and the closest-approach
distance in its partner are defined by
\begin{equation}
\sin ^2\varepsilon /2=F({\mathcal L,ZRZ}^{-1}),\quad
\sinh^2\delta/2=-F({\mathcal L}^{-1}{\mathcal ,ZRZ}^{-1}).
\end{equation}
Anticipating again  $\epsilon \approx \delta\ll 1$ we conclude
$F({\mathcal L,ZRZ}^{-1})\approx -F({\mathcal L}^{-1}{\mathcal
,ZRZ}^{-1})$ or, with the help of $\mathrm{Tr}{\mathcal
L}=\mathrm{Tr}{\mathcal L}^{-1} $, $\;\mathrm{Tr}{\mathcal
LZRZ}^{-1} \approx -\mathrm{Tr}{\mathcal L}^{-1}{\mathcal
ZRZ}^{-1}$. The approximate equality of the absolute values of the
two traces is clearly necessary for the length of the two orbits
in the pair to be close due to (\ref{1.1}); the minus sign is less
obvious.

\subsection{Crossing/avoided-crossing partnership}\label{partnership}

Even though we are mainly interested in  pairs of
orbits with small $\epsilon$ and $\delta$ it is interesting to
further dwell on the crossing/avoided-crossing partnership for
large angles and widths. We shall employ symbolic dynamics to
rigorously relate the crossing angle $\epsilon$ to the
closest-approach distance $\delta$ for such pairs and give exact
expressions for the action difference in a pair and  the
distribution $P^a(\delta|L)$. Such results might be of importance
for future investigations of higher-order terms in the expansion
of the form factor.

We again imagine the symbolic code of an orbit bisected as $ A B$;
then the partner orbit with respect to that bisection will have
the symbolic word $AB^{TR}$; the function $F({\cal A},{\cal B})$
defined in (\ref{myfo}) will become $F({\cal A},{\cal B}^{-1})$ for
the partner. As already indicated previously, numerical checks
revealed the following behavior of $F$ under this replacement:
(i) If the bisection $AB$ has $F({\cal A},{\cal B})>1$ the partner orbit $AB^{\rm TR}$
will also have $F({\cal A},{\cal B}^{-1})>1$;
(ii) $F({\cal A},{\cal B})<0\;$ entails $\; 0<F({\cal A},{\cal B}^{-1})<1$;
(iii) finally, if $\;0<F({\cal A},{\cal B})<1\;$ then, the partner orbit
has $F({\cal A},{\cal B}^{-1})<0$. If $A$ or $B$ contains less
than $5$ symbols, and $F$ very close to 1, very rare exceptions to
this rule occur.

Considering the previously established interpretations of the
three cases (\ref{aac}),\ (\ref{pac}), and (\ref{myfo}) we can say that
partial time inversion of an orbit has the following respective
consequences:
(i) A parallel avoided crossing is replaced by another parallel avoided crossing;
(ii) an antiparallel avoided crossing becomes a crossing in the partner;
(iii) a crossing becomes an antiparallel avoided crossing in the partner. For very
short loops with at most $5$ symbols
and angles close to $\pi$ very rare exceptions occur.

We can use the last of these findings to more precisely {\it
define} an antiparallel avoided crossing \footnote{From here on
all avoided crossing considered will be antiparallel such that the
qualifier ``antiparallel'' will be dropped.} as the object
emerging after the partial time reversal of some orbit $AB\to
AB^{\rm TR}$ \ (or, equivalently, $AB\to A^{\rm TR}B$ ) provided the bisection $AB$ has a crossing according
to $0<F({\cal A},{\cal B})<1$ while $F({\cal A},{\cal B}^{-1})<0$.
As always, the word $AB$ is assumed pruned in the sense of
Sect.~\ref{longorb}, i.e. written in the standard representation
such that each of its letters indicates a side of the fundamental
domain visited by the orbit. The code $A B^{\rm TR}$ of the
partner orbit with an avoided crossing will then, however, not
always be pruned, and the partner orbit will sometimes not have
all its circular sections within the fundamental domain.

An example of a bisection of a pruned word for which partial time
reversal yields a non-pruned word is provided by $A=LZ,\;B=RZ$.
Then  the symbolic word of the partner $AB^{\rm TR}=LZZ^{\rm
TR}R^{\rm TR}$ obviously contains mutually canceling parts, and
$LR^{TR}$ may become the standard encoding of an orbit in the
fundamental domain. However, the avoided-crossing width in this
case will be determined by the function $F({\cal L Z},{\cal
Z}^{-1}{\cal R})$ in whose arguments the insertion ${\cal Z}$ may
not be dropped. This situation occurs when the geodesic
determining the width of the avoided crossing  consists of several
segments when depicted inside the octagon (see the preceding
section).
There is no reason not to take into account avoided crossings of
the type just characterized.  An orbit with an avoided crossing
may thus contain fewer symbols and be appreciably shorter than its
partner with a crossing. The closer the crossing angle is to $\pi$
the more often one meets with that situation.

According to our definition there is a one-to-one correspondence
between crossings and avoided crossings. One might be tempted to
infer that in each long ergodic orbit the number of crossings must
be equal to the number of avoided crossings, but that temptation
must be resisted. Given a fixed length $L$, there are fewer
crossings than avoided crossings simply since by replacing a
crossing with an avoided crossing one gets a shorter periodic
orbit. Therefore, to obtain all avoided crossings with a certain
width $\delta$ in the periodic orbits of length $L$  one has to
make the crossing to  avoided-crossing replacement in all orbits
of length $L+\Delta L(\delta)$ with suitable length increments
$\Delta L$, and  longer orbits are more numerous due the
exponential proliferation.

To find the relation between the density $P(\epsilon|L)$ of
crossing angles and the density $P^a(\delta|L)$ of the
avoided-crossing widths we write the total number of crossings in
all periodic orbits with lengths in the interval $[L,L+dL]$ as
$N(L)P(\epsilon|L)dL d\epsilon $ with Huber's proliferation law
$N(L)=\mbox{e}^L/L$. On the other hand, the number of avoided
crossings with  width $\delta$ in these orbits will equal the
number of all crossings in the ``parent'' orbits whose lengths lie
in $[L+\Delta L,L+\Delta L +dL]$ which is $N(L+\Delta L)
P(\epsilon|L+\Delta L) dL |\frac{d\epsilon}{d\delta}| d\delta$. We
may thus write
\begin{equation}                                                  \label{denaco}
P^a(\delta|L)\approx \frac{\mbox{e}^{\Delta L}}{1+\frac{\Delta L}{L}} \
\left|\frac{d\epsilon}{d\delta}\right| \ P(\epsilon|L+\Delta L)\,,
\end{equation}
where the approximate-equality sign again signals large $L$. It
remains to determine the length shift $\Delta L(\delta)$ and the
crossing angle $\epsilon(\delta)$, both as functions of $\delta$.
Clearly, if we only wanted to find the ergodic part of
$P^a(\delta|L)$ from $P_{\rm erg}(\epsilon|L)= (L^2/2\pi
A)\sin\epsilon$ we could, in analogy with our procedure in
Sect.~\ref{ergorb}, drop the length shift $\Delta L$ everywhere in
(\ref{denaco}) except in the exponential proliferation factor;
actually, we must be more ambitious and go for the next-to-leading
order in $L$ as well since the latter is of relevance for the form
factor.

To determine the  functions $\mbox{e}^{\Delta L(\delta)}$ and
$\epsilon(\delta)$ entering the foregoing expression we invoke two
consequences of the definition (\ref{su11matrix}) of $SU(1,1)$
matrices ${\cal A,B}$. First, we note $ {\cal B}+ {\cal
B}^{-1}=\mathbf{1} \mbox{Tr }\cal{B}$ where $\mathbf{1}$ is the
unit matrix. Second, on multiplying with $\cal A$ and taking the
trace we get
\begin{equation}
\label{matrid}
 \mbox{Tr}{\cal AB} + \mbox{Tr}{\cal AB}^{-1}=\mbox{Tr}{\cal A} \mbox{ Tr}\cal{B}\,.
\end{equation}
Using the identity (\ref{matrid}) we can rewrite the expressions
(\ref{myfo}) for the crossing angle and (\ref{aac}) for the
avoided-crossing width as
\begin{eqnarray}
\label{epsdel} \cos^2  \frac{\epsilon}{2}= \frac{ (\mbox{Tr}{\cal
A})^2+(\mbox{Tr}{\cal A})^2-4 - \mbox{Tr}{\cal AB} \mbox{ Tr}
{\cal AB}^{-1}}
 {(\mbox{Tr} {\cal AB})^2-4 } \nonumber\\
 \cosh^2\frac{\delta}{2}= \frac{(\mbox{Tr}{\cal A})^2+ (\mbox{Tr}{\cal A})^2-4
-\mbox{Tr}{\cal AB}\mbox{ Tr}{\cal AB}^{-1}} {(\mbox{Tr}{\cal
AB}^{-1})^2-4 }
\end{eqnarray}
and conclude
\begin{equation}
\frac{\cos^2  \frac{\epsilon}{2}}{ \cosh^2\frac{\delta}{2}}=
\frac{(\mbox{Tr} {\cal AB}^{-1})^2-4}{(\mbox{Tr}{\cal AB})^2-4}\,.
\end{equation}
Herein employing (\ref{1.1}) and denoting the lengths of the periodic
orbits with crossing and avoided crossing by $L_\epsilon$ and
$L_\delta$ we proceed to
\begin{equation}                                                  \label{lengths}
 \frac{\cos \frac{\epsilon}{2} }{\cosh  \frac{\delta}{2} }=
 \frac{\sinh  \frac{L_\delta}{2} }{\sinh \frac{L_\epsilon}{2} }\,.
\end{equation}

As by now familiar, slight simplifications arise in the physically
interesting case of loops long enough for $\mbox{Tr} {\cal AB}$ to
be much larger than both $\mbox{Tr}{\cal A}$ and $ \mbox{Tr}{\cal
B}$. The rigorous symbolic-dynamics expressions  (\ref{epsdel}) for
$\epsilon$ and $\delta$ then simplify to
\begin{equation} \label{approco}
\cos^2\frac{\epsilon}{2}\approx - \frac{\mbox{Tr } {\cal
AB}^{-1}}{\mbox{Tr } {\cal AB}}\,,\;\;\;\;
 \cosh^2  \frac{\delta}{2}
\approx - \frac{\mbox{Tr } {\cal AB}}{\mbox{Tr } {\cal AB}^{-1}}
\end{equation}
and yield the previously announced relation between the crossing
angle in the parent orbit  and the avoided crossing width in the
partner,
\begin{equation}
\label{stefrel} \cosh \frac{\delta}{2}\approx \frac{1}{\cos
\frac{\epsilon}{2} }\,,
\end{equation}
which implies the familiar $\epsilon\approx\delta$ for small
angles. Similarly, we may combine (\ref{lengths}) and (\ref{approco}) to
\begin{equation}                                                 
- \frac{\mbox{Tr} {\cal AB}^{-1}}{\mbox{Tr} {\cal AB}}=
\frac{\cosh  \frac{L_\delta}{2} }{\cosh  \frac{L_\epsilon}{2} }
\approx\mbox{e}^{ \frac{L_\delta-L_\epsilon}{2} }=\mbox{e}^{- \frac{\Delta
L}{2} }
\end{equation}
and thus get the length (and thus action) difference for an orbit
with a self-crossing and its partner avoiding that crossing,
\begin{equation} \label{deltact} \Delta L\approx -4 \ln \cos  \frac{\epsilon}{2}= 4
\ln \cosh  \frac{\delta}{2}\,;\label{epsdelDel}
\end{equation}
the latter relation generalizes the  result of Sieber and Richter \cite{gedisieb,siebrich},
$\Delta L=\epsilon^2/2$, to all orders in $\epsilon$.

An interesting geometric interpretation of our relations
(\ref{epsdelDel}) between $\epsilon,\delta$, and
$\Delta L$ in the framework of hyperbolic geometry will be
presented in Appendix \ref{pseudeuc}. It may be well to once more
say that the approximate-equality sign in these relations points
to an error  which is  exponentially small when the loop lengths
are large.

We can now write the factors converting the density of crossing
angles into the density of avoided-crossing widths in (\ref{denaco}),
\begin{equation}                                                  \label{conversion}
\mbox{e}^{\Delta L}= \frac{1}{\cos^4  \frac{\epsilon}{2} }, \quad
 \frac{d\epsilon}{d\delta}= \frac{\tanh  \frac{\delta}{2} }
{ \tanh\frac{\epsilon}{2} }\,.
\end{equation}
The latter identities yield ${\rm e}^{\Delta
L}|\frac{d\epsilon}{d\delta}|\,\sin\epsilon =\sinh\delta$ such that
the ergodic angle distribution  $P_{\rm erg}(\epsilon|L)=(L^2/2\pi
A)\sin \epsilon$ yields the ergodic distribution of
avoided-crossing widths $P_{\rm erg}^a(\delta|L) = (L^2/2\pi A)
\sinh \delta$. Indulging in higher ambitions, we retain all terms
of the next-to-leading (i.e. first) order w.r.t. to $L$
contributed by $(1+\frac{\Delta L}{L})^{-1}\,P(\epsilon|L+\Delta L)$
in (\ref{denaco}). Taking the full angle density (\ref{lout},\ref{leffprox})
\begin{equation}
P(\epsilon|L)=\frac{L}{\pi A}\sin\epsilon\, \left(\frac{L}{2}
-2\ln\frac{c}{2\sin\frac{\epsilon}{2}}\right)
\end{equation}
from Sect.~\ref{joint} and using the above identities
(\ref{conversion}) to check $\Delta L-4\ln
(c/2\sin\frac{\epsilon}{2})=-4\ln (c/2\sinh\frac{\delta}{2})$ we get
\begin{equation}                                                  \label{ergdist}
P^a(\delta|L) = \frac{L}{\pi A} \sinh \delta\,
\left(\frac{L}{2}-2\ln\frac{c}{2\sinh\frac{\delta}{2}}\right)\,.
\end{equation}
As a most welcome surprise,
after combining all the ostensibly unrelated
 factors in (\ref{denaco}), including those resulting from Huber's
exponential proliferation law, the final result is simply the
analytical continuation of the crossing distribution to imaginary
crossing angles.

\subsection{Example of a Sieber-Richter pair}\label{expair}

\begin{figure}
\begin{center}
\leavevmode \epsfxsize=0.95 \textwidth \epsffile{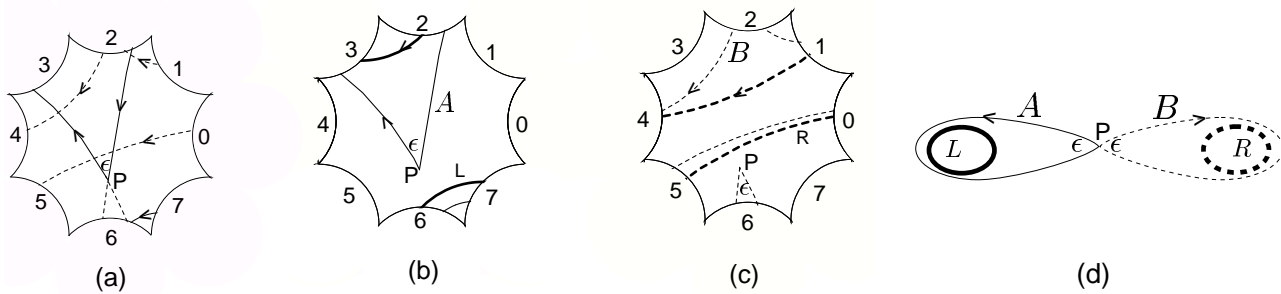}
\end{center}
\caption{(a): Loops $A=3, 6$ (full  line) and $B=6, 4, 5, 2$ (dashed) make
up  periodic orbit $ A B = \{ 3, 6, 6, 4, 5, 2 \} $. (b): Periodic
orbit  L = $\{ 3, 6 \} $ (thick line) and loop A. (c):
Periodic orbit  $R=\{ 4, 5 \} $ (thick dashed line)
and loop B. (d): Schematic picture of orbits $L, R$ and longer orbit $A B = \{ L Z R Z^{\rm TR} \} $ with crossing at point P obtained by deformation
of $L$ and $R$.} \label{zmz2}
\end{figure}

In the previous subsection we have already commented on the symbolic codes for the periodic orbits in a Sieber-Richter pair as $\{AB\}$ and
$\{A^{\rm TR}B\}$. To further clarify the role of symbolic dynamics for crossings, avoided crossings, and Sieber-Richter pairs  we would  like to present a concrete
such pair in the regular octagon. Fig.~\ref{zmz2} (a)
displays a numerically found periodic orbit with a small-angle crossing. The reader is invited to  read off the orbit segments, starting e.g. at the side 6  in the
direction shown by arrows obtaining  $ 6\to 3,\;7\rightarrow
6,\;2\rightarrow 6,\;2\rightarrow
4,\;0\rightarrow 5,\;1\rightarrow 2$ thus closing the loop. The  starting side is always opposite to the landing side of the preceding segment, with their numbers differing by $\pm4$. Hence it is sufficient to list the consecutive landing sides only: $\{\underline{3},6,\underline{6},4,5,2\}$. The sequence we got   is  the standard symbolic code of the orbit; the underlined symbols indicate the pair of segments crossing at the point $P$.

The crossing $P$ divides the orbit into two loops shown by the
full and dashed lines in Fig.\ref{zmz2} (a); the symbol sequences
of both loops $A= 3,6$ and $B= 6,4,5,2$ start with the symbol of
the segment participating in the crossing. The periodic orbits
associated with the loops and the loops themselves are shown by
thick and thin lines respectively in Fig.~\ref{zmz2}(b) for loop
$A$ and (c) for loop $B$. The periodic orbit $\{L\}$ associated
with the loop $A$ has the same symbol sequence $\{3,6\}$ as the loop
itself. The code of the orbit associated with the loop $B$ is
shorter than that of the loop, namely, $\{R\}= \{ 4,5 \}$ while
$B=ZRZ^{{\rm TR}}$ with $Z=6$; of course, the insertions $Z$ fall out
of the code of the loop-associated orbit. Inspection of part
  of Fig.~\ref{zmz2}(c) helps to appreciate insertions  $Z$
in a loop like $B=ZRZ^{\rm TR}$ as due to the deformation of a
circular segment of the periodic orbit $\{R\}$ overstepping the boundary of the octagon.  Speaking pictorially we may say that the middle of the upper segment of $R$ was dragged upwards such
that the  vertex of the loop crossed the boundary 2 of the octagon,
reappeared at the opposite boundary 6, and finally became the crossing point
$P$, hence the insertions  $Z=6$ and $Z^{TR}=2$. On the other hand, when deforming segment 3 of $L=3,6$ to the loop
$A$ ( Fig.~\ref{zmz2}(c)) the octagon boundary is never reached, and therefore no new
symbols appeared in the loop $A$ compared with the periodic orbit
$\{A\}$.

\begin{figure}
\begin{center}
\leavevmode
\epsfxsize=0.44
\textwidth
\epsffile{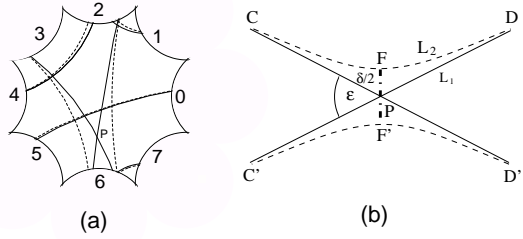}
\end{center}
\caption{ (a) Orbit of Fig.~\ref{zmz2} with crossing at $P$ and partner avoiding that crossing.  (b) Definition of crossing angle $\epsilon$ and avoided-crossing width $\delta$. When crossing/avoided crossing geometry (b) is folded into the fundamental domain $C$ is identified with  $C'$ and $D$ with $D'$, and representation (a) results.}
\label{cravcr}
\end{figure}

To obtain the Sieber-Richter partner of the orbit $\{AB\}$ in Fig. \ref{zmz2}
(a) we have to replace one of its loops, e.g. $A=3,6$  by its time reverse
$A^{\rm TR}=\overline{6},\overline{3}=2,7$ . The partner periodic
orbit thus formed $\{2,7,6,4,5,2 \}$ is shown in Fig \ref{cravcr}
(a) (dotted line) together with the original orbit (full line).
The partner has an antiparallel avoided crossing  whose width is
approximately equal to the crossing angle in the original orbit;
the relation valid to all orders in $\epsilon$ is given by
(\ref{stefrel}). Note that apart from the two segments participating in
the crossing (resp. avoided crossing)  the two orbits run very close
together; thus it is these two segments  which contribute mainly to the
difference in the orbit lengths.

\vspace{1cm} Financial support by the Sonderforschungsbereich
"Unordnung und gro{\ss}e Fluktuationen" of the Deutsche
Forschungsgemeinschaft is gratefully acknowledged. We have
profited from discussions with Uwe Abresch, Ralf Aurich, Predrag
Cvitanovi\'c, Gerhard Knieper, Christopher Manderfeld, Klaus Richter, Martin Sieber, Uzy Smilansky, Dominique Spehner, Frank Steiner, and Wen-ge Wang.

\appendix
\section{Action difference for  pairs from hyperbolic triangles}\label{pseudeuc}

Consider a triangle $XYZ$ formed by geodesics on the pseudosphere, with the
angles $X,Y,Z$  named after the respective vertices  and the lengths of the respective opposite sides $x,y,z$.
The
following generalizations of elementary Euclidian-geometry
relations hold true \cite{geobo}
\begin{eqnarray}
\cosh x = \frac{\cos Y \cos Z + \cos X}{\sin Y \sin
Z}, \quad
 \frac{\sin X}{\sinh x}= \frac{\sin
Y}{\sinh y} = \frac{\sin Z}{\sinh z} \,. \label{geq2}
\end{eqnarray}

Let us apply these identities to the triangle $FPD$ in Fig.~\ref{cravcr} (b) depicting the crossing and avoided crossing  in a Sieber-Richter pair.
Again naming the angles of the triangle after the respective vertices we have
$F=\pi/2,P=\pi/2-\epsilon/2$. The point $D$ is supposed to be  removed to infinity (recall the assumption of long
orbits) such that the lengths $FD=L_2$ and $PD=L_1$ tend to infinity while the angle $D$ tends to zero. However, the difference $L_2-L_1$ remains finite
 and gives, up to an
obvious factor 4, the length (or action) difference in search,
$\Delta L= L_{\epsilon} - L_{\delta}=4(L_1-L_2)$. From the first of the relations (\ref{geq2}) we
get the connection between the crossing angle $\epsilon$ and the
closest-approach distance $\delta=2 FP$,
\begin{equation}
\cos\frac{\epsilon(\delta)}{2} \approx \frac{1}{\cosh\frac{\delta}{2}}
\label{deltay}\,,
\end{equation}
which we recognize as (\ref{stefrel}); similarly, the second group of relations  in (\ref{geq2}) yields
the length difference
\begin{equation}
\cos \frac{\epsilon}{2} = \frac{\sinh L_1}{\sinh L_2} \approx e^{L_1 -
L_2} \;\Longrightarrow\;\Delta L\approx 4\ln\cos\frac{\epsilon}{2}
\label{yea}
\end{equation}
in agreement with (\ref{deltact}).
The association of symbolic dynamics and  M\"obius-transformation
matrices employed in Sect.~\ref{symbdyn} to derive the foregoing
relations obviously incorporates the hyperbolic geometry of the
Hadamard-Gutzwiller model.

\section{Effective number of symbols}\label{peffA}
We here determine the number $\nu (n)$ of non-equivalent $n$-letter
words (alias  periodic orbits
with $n$ segments) $\gamma$    with
an alphabet containing $p=8$ different letters. As mentioned in Section
V, a rough estimate is $\nu(n) \approx (p-1)^n/n$, since a symbol
$j$ may not be succeeded by its ``inverse'' $\bar{j}=(j+4){\rm\mbox{ }mod\mbox{ }}8$, and since a factor $1/n$ is due to the identification of
cyclic permutations.  The influence of the group identity
\begin{eqnarray}
(0, 5, 2, 7, 4, 1, 6, 3) = \mathbf{1} \,. \label{all}
\end{eqnarray}
requires more thought.

Our first task is to find the shortest possible version of a given
word. Toward that end we note that whenever the sequence
(\ref{all}) or any of its cyclic permutations is encountered
within a word, we must delete that sequence and thus shorten the
word  by $8$ letters. Similarly, if some part of the identity with
$m$ symbols and $4<m \leq 8$, is encountered, that part must be
replaced by its complement with $8-m$ symbols. In any such
instance the word is shortened by 8,6,4 or 2 letters.
Next, we have to consider the  $4$-symbol sequences for which the group
identity provides an equivalent $4$-symbol sequence like for
example $(0, 5, 2, 7) = (7, 2, 5, 0)$.
If a word contains $k$ such 4-letter patches there are $2^k$
different representations of the same orbit; only one of these
representations has the property that each symbol corresponds to
one segment of the orbit inside the fundamental domain.

Let us start with the reversible 4-letter sequences and denote by $\alpha$ the probability that a symbol of a long word is  the beginning of such a  sequence. A combinatorial estimate of $\alpha$
can be made as follows. There are $16$ different $4$-letter
patterns of the type in question, due to 8 cyclic permutations of
both the identity in (\ref{all}) and  its inverse,  and each of
these patterns occurs with probability $1/(8\times7^3)$. However, the
pattern can only be included when the first and last letter are
not the inverse of the neighboring one in the complete word, such
that a "junction factor" $(7/8)^2$ must be incorporated. The estimate
in search thus comes out as $\alpha_{\rm comb} = (7/8)^2\times 16/(8 \times 7^3)
\approx 0.00446$. Assuming independent occurrence of
the 4-letter sequences we expect to find exactly $k$ of them  in an
$n$-letter word with the binomial probability
\begin{eqnarray}
P_n (k, \alpha) =  \frac{n!}{k! (n-k)!} \alpha^k (1 - \alpha)^{n -
k}\,. \label{binalpha}
\end{eqnarray}
We refrain from attempts at improving that
combinatorial estimate by accounting for the ``interaction'' of the
4-letter patches caused by their finite length. Upon checking many long
words with randomly chosen letters we found such interaction effects
quite unimportant: The binomial distribution (\ref{binalpha}) is borne
out very well with $\alpha_{\rm num}=0.00425$.

Similarly, let $\beta$ be the probability to find, in any letter,
the beginning of any of the previously discussed $m$-symbol
patterns with $4<m\leq 8$ which has an equivalent shorter partner
with $8-m$ symbols. Again assuming independence of several such
events we get a binomial distribution $P_n(k,\beta)$ as before and
the combinatorial estimate for $\beta$ is, again neglecting
``interactions'', $\beta_{\rm theor} = (7/8)^2 [16/(7\times 8^7)+ 16/(7
\times8^6) + 16/(7 \times8^5) + 16/(7\times 8^4)] = 0.00074$. Here again, the
interactions were checked to be unimportant by going through a
large sample of random words. The numerical data for $P_n^{\rm
num}(k)$ so obtained once more fit the binomial distribution, with
$\beta_{\rm num} = 0.000716$.
.

We can finally put all pieces together. The number of $n$-letter
words allowed after excluding $j{\bar j}$ sequences of symbols and
accounting for the equivalence of cyclic permutations was seen to
be $(p-1)^n/n$; only the fraction $(1-\beta)^n$ of these cannot be
shortened by using the group identity, so we are down to
$(1-\beta^n)(p-1)^n/n$ words. Roughly
$P_n(k,\alpha)(1-\beta)^n(p-1)^n/n$ of these, however, contain
precisely $k$ sequences of four letters each of which comes in two
equivalent pairs due to the group identity; to avoid overcounting
we have to divide out the multiplicity $2^k$. Finally summing over
$k$ we get the desired number of different $n$-letter words as
\begin{eqnarray}
\nu (n)&=& \sum_k \frac{n!}{k! (n-k)!} \frac{1}{2^k} \alpha^k (1 -
\alpha)^{n - k} (1 - \beta)^n (p-1)^n/n \nonumber\\ &=&
\frac{1}{n} [(1 - \alpha/2)(1 - \beta)(p-1)]^n  = \frac{1}{n}
{\rm e}^{n \ln p_{\rm eff} }\,,\\ p_{\rm eff} &=& (1 - \alpha/2)(1 -
\beta)(p-1) = 6.98\,. \label{peff}
\end{eqnarray}
Interestingly, the effective number of symbols is not drastically
reduced from $p=8-1=7$ by the group identity equation (\ref{ide}).

\end{document}